\documentclass[%
 aip,
 amsmath,amssymb,
 reprint,%
]{revtex4-1}

\usepackage{graphicx}
\usepackage{dcolumn}
\usepackage{bm}

\usepackage[utf8]{inputenc}
\usepackage[T1]{fontenc}
\usepackage{mathptmx}
\usepackage{etoolbox}
\usepackage{color}
\usepackage{epstopdf, epsfig}
\usepackage{overpic}
\usepackage{enumerate}
\usepackage[dvipsnames]{xcolor}
\usepackage[colorlinks,
            linkcolor=blue,
            anchorcolor=blue,
            citecolor=blue]{hyperref}

\usepackage{makecell}
\usepackage{amssymb} 

\usepackage{mathrsfs}

\usepackage{threeparttable}
\usepackage{tabularx}
\usepackage{CJKutf8}

\makeatletter
\def\@email#1#2{%
 \endgroup
 \patchcmd{\titleblock@produce}
  {\frontmatter@RRAPformat}
  {\frontmatter@RRAPformat{\produce@RRAP{*#1\href{mailto:#2}{#2}}}\frontmatter@RRAPformat}
  {}{}
}%
\makeatother
\begin{document}

\preprint{AIP/123-QED}

\title[Sample title]{On the relationships between pressure geometrization and vortex identification}
\author{Han Tu \begin{CJK*}{UTF8}{gbsn}(涂晗)\end{CJK*}}
\affiliation{Hubei Provincial Key Laboratory of Chemical Equipment Intensification and Intrinsic Safety, School of Mechanical and Electrical Engineering, Wuhan Institute of Technology, Wuhan 430205, China}

\author{Hang Zhao \begin{CJK*}{UTF8}{gbsn}(赵航)\end{CJK*}}
\affiliation{AECC Hunan Aviation Powerplant Research Institute, Zhuzhou 412002, China}

\author{Qi Gao \begin{CJK*}{UTF8}{gbsn}(高琪)\end{CJK*}}
\email{qigao@zju.edu.cn}
\affiliation{School of Aeronautics and Astronautics, Zhejiang University, Hangzhou 310027, China}%

\date{\today}

\begin{abstract}
The profound mathematical and conceptual analogy between pressure in fluid mechanics and gravity in general relativity motivates a novel geometric reinterpretation of flows.
This article proposes a theoretical framework for geometrizing pressure by establishing a Newton–Cartan geometry and applying a conformal transformation to the spatial metric. 
The pressure field is reinterpreted as a geometric potential that induces an effective spatial curvature, encoded in the Ricci tensor $\tilde{R}_{ij}$ and the associated Einstein-like tensor $\tilde{G}_{ij}$.
Within this curved effective space, fluid particles under the exclusive influence of pressure follow geodesics, and the classical Euler equation is recovered in the weak-field limit. The pressure Poisson equation emerges naturally as a consequence of the geodesic equation and the Bianchi identity, revealing pressure as an instantaneous geometric constraint rather than an external force. 
This framework provides a unified geometric interpretation of widely used vortex identification criteria, demonstrating that the prevailing Eulerian and Lagrangian schemes correspond to different geometric invariants or projections of the effective curvature. 
In the degenerate two-dimensional (2D) limit, the effective spatial tensor vanishes identically, and the global integral of the total curvature vanishes as well. 
This topological constraint offers a novel interpretation of the dual-cascade scenario in 2D turbulence: the inverse energy cascade arises from geometric condensation of positive-curvature structures toward larger scales, while the forward enstrophy cascade results from geometric dispersion of negative-curvature structures toward smaller scales.

\end{abstract}

\maketitle

\section{Introduction}\label{Intro}

The profound similarity between pressure in fluid mechanics and gravity in general relativity can be noticed at first glance from the Poisson equations for the pressure potential ($\phi_{p}$) and the gravity potential ($\phi_{g}$):
\begin{equation}
\begin{aligned}
\nabla^{2}\phi_{p}&=-\partial_{j}u^{i}\partial_{i}u^{j},\\
\nabla^{2}\phi_{g}&=4\pi\rho.
	\label{equ1}
\end{aligned}
\end{equation}
Here $\partial_{j}u^{i}$ and $\rho$ represent the velocity gradient and mass density, respectively, serving as the source terms that determine the distributions of both fields.
This formal analogy, however, is merely the surface of a deeper structural parallel.
Einstein's monumental achievement was to elevate gravity from a force to a manifestation of spacetime curvature, positing that a particle in free fall follows a geodesic \cite{Einstein1922}. Analogously, a fluid particle, under the exclusive influence of pressure, follows a streamline prescribed by the Euler equation.
Beyond these mathematical parallels, the striking morphological resemblance between black holes and vortices further inspires the quest for a framework that geometrizes pressure.

The idea of analogue gravity in fluids dates back to the seminal work of \citet{unruh_experimental_1981}, which employs the velocity field to construct an effective metric capable of generating curvature. This approach has since found wide application in acoustic studies (e.g., acoustic black holes) \cite{rousseaux_observation_2008,rousseaux_horizon_2010,braunstein_analogue_2023}.
For the pressure–gravity analogy, however, it is more natural to ascribe the role of curvature to the pressure field itself. Accordingly, pressure should be interpreted not as an external force but as an intrinsic manifestation of spatial curvature, thereby establishing a direct geometric correspondence.

The central challenge in realizing this geometrization lies in the foundational structure of non-relativistic fluid mechanics.
Unlike general relativity, where curvature resides in spacetime itself, incompressible flows evolve within an absolute time and a flat Euclidean space, which appears to offer no natural stage for curvature to exist.
A proper approach would be to apply a conformal transformation utilizing an effective metric, a technique that has been widely used for airfoil shape generation as well as in acoustics \cite{unruh_experimental_1981,barcelo_analogue_2005}.
By conformally mapping the physical space, an effective space in which the influence of pressure is encoded as a geometric property can be constructed, thus providing a fertile ground for the emergence of curvature.

As black holes curve spacetime and stand as the most profound objects in cosmology, vortices exhibit distinct geometry and vigorous motion, attracting researchers from different branches of fluid mechanics \cite{haller2013coherent,xiong_identifying_2019,qian_vortex_2022}.
Essentially, vortices could be viewed as the genes of fluid motions \cite{wang_novel_2022}, and they are regarded as the building blocks of turbulent motions \cite{lundgren_strained_1982,adrian_hairpin_2007}.
In our current framework, vortices are expected to induce curvature in the effective space much as black holes do in spacetime. Consequently, the identification of vortices and the quantification of their dynamic characteristics provide indispensable information for understanding complex flows.


Due to the ambiguity of the vortex definition and the difficulty of mathematically describing such coherent structures, a well-accepted identification scheme for a vortex is not yet available \cite{epps_review_2017,wang_novel_2022,wang_effectiveness_2022}.
Various identification criteria have been proposed to recognize vortical structures from velocity fields. 
Specifically, most of these criteria are calculated from the velocity gradient tensor (VGT) and shape local kinematic characteristics by describing streamlines or the rotation/strain tensor, such as the second invariant $Q$ \cite{moin_eddies_1988} , the swirl strength $\lambda_{ci}$ \cite{zhou_mechanisms_1999} and the $\lambda_{2}$ criterion \cite{Jeong_Hussain_1995}.
By defining scalar fields, these criteria reflect the local vortical intensity.
In addition to these criteria that characterize vortex intensity, there are also several schemes that extract the vortex orientation.
A typical approach is to use the real eigenvector $\Lambda_{r}$ of the VGT to define the vortex orientation \cite{gao_analysis_2011,tian_vortex_2020}. $\Lambda_{r}$ has been demonstrated to align well with the vortex centerline \cite{zhou_self-sustaining_1997} and to perform best with respect to the effective representation of wall turbulence \cite{wang_effectiveness_2022}.

However, it is crucial to recognize that these criteria depend exclusively on the VGT, that is, on kinematic information alone \cite{wang_novel_2022}.
As a consequence, dynamic information (i.e., pressure and its derivatives) is neither taken into account nor implicitly embedded.
A direct interpretation of these criteria from the pressure side is therefore necessary to deepen our dynamical understanding, not only of the criteria themselves but also of the very nature of vortices.
Although the idea of general relativity has been widely applied in the fluid community (e.g., relativistic fluid mechanics \cite{rieutord_introduction_2006,weickgenannt2022newdevelopmentsrelativistichydrodynamics,berbel_exact_2023,jain_schwinger-keldysh_2024}), and the geometrization of fluids has been vigorously developed (e.g., \citet{arnold1966geometrie} proved that the Euler equation can be regarded as the geodesic flow on the group of volume-preserving diffeomorphisms SDiff with respect to the $L^{2}$ metric; hydrodynamics on curved surfaces \cite{armas2020newton} has been developed for studies of fluid membranes in biological systems \cite{canham1970minimum,helfrich1973elastic} and surface/edge physics in condensed matter systems \cite{kane2005z}), to the best knowledge of the authors, very little literature has reported the establishment of a pressure geometrization framework for non-relativistic flows.




Therefore, the current work aims to render this implicit connection explicit and geometric by offering a novel geometric perspective on pressure, inspired by the profound analogy to gravity.
The pressure itself, the pressure Poisson equation, and the classical Euler equation are carefully revisited and furnished with geometric interpretations. 
Moreover, a direct reinterpretation of current vortex identification schemes is offered in terms of pressure geometrization.
This framework not only elucidates the role of pressure as a geometric potential but also unifies a variety of vortex criteria under a common geometric language, thereby enriching our understanding of coherent structures in fluid flows.

This article is organized as follows. 
In Sect.~\ref{con}, the detailed methodology of the geometrization procedure and its corresponding physical interpretations are presented.
Section~\ref{con:2D} discusses the geometrization characteristics specific to degenerate two-dimensional (2D) flows, with particular emphasis on the influence of dimensional reduction on vortical structures.
In Sect.~\ref{pre_vid}, the geometric scope of current vortex identification schemes is provided.
We end this article with discussions in Sect.~\ref{Discussions}.

\section{Methodology of geometrization and physical interpretation}\label{con}

\subsection{Establishment of Newton–Cartan geometry and geometrization of pressure}\label{con:NC}
We consider the motion of an inviscid fluid in flat Euclidean space. 
To geometrise the pressure gradient, we model the spacetime as a Newton–Cartan (NC) geometry \cite{andringa_newtonian_2010,hansen_action_2019}, which is the natural framework for describing non-relativistic gravity and fluids. 
The NC geometry consists of the fields $(\tau_{\mu},h_{\mu \nu})$, where the Greek indices denote spacetime indices such that $\mu, \nu=0,1,2,3$.
It is useful to define an inverse NC structure $(v^{\mu},h^{\mu \nu})$, where $v^{\mu}$ spans the kernel of $h_{\mu\nu}$ and $\tau_{\mu}$ spans the kernel of $h^{\mu\nu}$.
Specifically, the 1-form $\tau_{\mu}$ is also called the $clock$ $1$-$form$, as expressed below:
\begin{equation}
    \tau_{\mu}=\partial_{\mu}t=(1,0,0,0).
    \label{con:NC_eqn1}
\end{equation}
This ensures the absoluteness of time (i.e., no expansion or compression in time), with physical time intervals given by $\Delta T = \int \boldsymbol{\tau}$. The vector $v^{\mu}$ is known as the NC velocity.

The space metric tensor $h_{\mu \nu}$ is a degenerate $(2,0)$ type tensor, given by
\begin{equation}
    h^{\mu\nu}=
\begin{pmatrix}
0 & 0 \\
0 & \delta^{ij}
\end{pmatrix}.
    \label{con:NC_eqn2}
\end{equation}
Thus, a degenerate spatial metric is given by $\boldsymbol{h} = h^{\mu\nu} \partial_\mu \otimes \partial_\nu$, satisfying the orthogonality condition $h^{\mu\nu}\tau_\nu = 0$ and ensuring that the curvature caused by the pressure gradient occurs only in space. Introducing spatial coordinates $x^i$ ($i=1,2,3$), the background spatial metric reduces to the Euclidean metric $\delta^{ij}$.
Additionally, these structures satisfy the completeness relation and normalization condition:
\begin{equation}
\begin{aligned}
\delta_\nu^\mu&=-v^\mu\tau_\nu+h^{\mu\rho}h_{\rho\nu},\\
v^{\mu}\tau_{\mu}&=-1.
	\label{con:NC_eqn2_1}
\end{aligned}
\end{equation}
The restricted setting of a flat NC background (i.e., a spacetime with absolute time whose constant time slices are described by the flat Euclidean geometry) is the most relevant case in the context of hydrodynamics on curved surfaces. More details of the necessary geometric tools to define an NC structure can be found in \citet{armas_newton-cartan_2020}.

To geometrise the pressure gradient $\nabla p$ as a manifestation of spatial curvature, we introduce an effective spatial metric $\tilde{h}^{ij}$ that depends on the pressure field. We postulate a conformal transformation
\begin{equation}
\tilde{h}^{ij}=\mathrm{e}^{-2\phi}h^{ij}=\mathrm{e}^{-2\phi}\delta^{ij},
    \label{con:NC_eqn3}
\end{equation}
where the scalar potential $\phi$ is defined as:
\begin{equation}
\phi=\frac{p}{\rho c_0^2}.
    \label{con:NC_eqn4}
\end{equation}
Here $c_0$ is the characteristic velocity (e.g., the speed of sound), introduced to maintain correct dimensions.
We adopt natural units with the sound speed set to unity, $c_{0}=1$, so that the dimensionless pressure potential reduces to $\phi=p/\rho$.
The corresponding covariant spatial metric is $\tilde{h}_{ij} = \mathrm{e}^{2\phi} \delta_{ij}$ (see Appendix \ref{A1}).
The physical interpretation of this construction is that regions of higher pressure correspond to a metric dilation, i.e., space is ``expanded'' due to a larger scale factor of the effective spatial metric, whereas low-pressure regions correspond to metric contraction, i.e. space is ``compressed''.



\subsection{Interpretation of spatial connection and curvature}\label{con:Rij}

The effective metric $\tilde{h}^{ij}$ induces a spatial connection $\tilde{\Gamma}^{i}_{jk}$. From the metric compatibility condition $\tilde{\nabla}_{k} \tilde{h}^{ij} = 0$, the connection coefficients are
\begin{equation}
\tilde{\Gamma}_{jk}^{i}=\frac{1}{2}\tilde{h}^{il}\left(\partial_{j}\tilde{h}_{kl}+\partial_{k}\tilde{h}_{jl}-\partial_{l}\tilde{h}_{jk}\right).
    \label{con:NC_eqn5}
\end{equation}
Substituting $\tilde{h}_{ij} = \mathrm{e}^{2\phi} \delta_{ij}$ yields the explicit expression (see Appendix \ref{A2})
\begin{equation}
\tilde{\Gamma}_{jk}^i=\delta_j^i\partial_k\phi+\delta_k^i\partial_j\phi-\delta_{jk}\partial^i\phi,
    \label{con:NC_eqn6}
\end{equation}
where $\partial^i \phi = \delta^{il} \partial_l \phi$. 
Thus the pressure information directly enters the spatial connection.

From the connection we compute the spatial Ricci tensor $\tilde{R}_{ij}$. For a three-dimensional (3D) conformally flat metric (i.e., a spacetime with absolute time), the standard formula gives
\begin{equation}
\tilde{R}_{ij}=-\partial_i\partial_j\phi-\delta_{ij}\nabla^2\phi+\partial_i\phi\partial_j\phi-\delta_{ij}(\nabla\phi)^2,
    \label{con:NC_eqn7}
\end{equation}
where $\nabla^2 = \delta^{ij}\partial_i \partial_j$ is the Laplacian of the flat background. 
Notably, the second order terms of the Ricci tensor are represented by $\partial_i\partial_j\phi$ and $\delta_{ij}\nabla^2\phi$, which indicate a significant contribution of the pressure Hessian and the pressure Laplacian to the curvature.
The corresponding scalar curvature can be calculated as:
\begin{equation}
\tilde{R}=\tilde{h}^{ij}\tilde{R}_{ij}=\mathrm{e}^{-2\phi}\left[-4\nabla^2\phi-2(\nabla\phi)^2\right].
    \label{con:NC_eqn8}
\end{equation}
Following the format of the Einstein tensor \cite{Einstein1922}, we further define an effective spatial tensor $\tilde{G}_{ij} = \tilde{R}_{ij} - \tfrac{1}{2} \tilde{R} \tilde{h}_{ij}$, given as:
\begin{equation}
\tilde{G}_{ij}=-\partial_i\partial_j\phi+\partial_i\phi\partial_j\phi+\delta_{ij}\nabla^2\phi.
    \label{con:NC_eqn9}
\end{equation}
In the weak-field limit $|\phi| \ll 1$ (e.g., incompressible flows with $p/(\rho c_{0}^2)\ll 1$), we have $\tilde{G}_{ij} \simeq -\partial_i \partial_j \phi + \delta_{ij} \nabla^2 \phi$, and its trace becomes $\tilde{G} = \tilde{h}^{ij} \tilde{G}_{ij} \simeq 2\nabla^2 \phi$.
Similar to the Ricci tensor, the spatial tensor $\tilde{G}_{ij}$ is also dominated by the pressure Hessian and Laplacian terms.
The detailed derivation process for these two tensors and their corresponding traces is provided in Appendix~\ref{A3}.

\subsection{Equivalence of geodesics and Euler equation}\label{con: geodesics}

The central interpretation of the present work is that the pressure gradient can be geometrised as an effective spatial curvature.
Analogously to the Einstein equations, the pressure term, which resembles the gravitational term in general relativity, is absorbed into the spatial geometric properties and no longer manifests as an external force in the equation of motion for a fluid particle.
Consequently, the four-acceleration of a fluid particle must vanish, and the particle follows a geodesic in the effective space:
\begin{equation}
a^{\nu}=u^{\mu}\nabla_{\mu}u^{\nu}=\frac{\mathrm{d}u^{\nu}}{\mathrm{d}t}+\tilde{\Gamma}^{\nu}_{\alpha\beta}u^{\alpha}u^{\beta}=0.
    \label{con:NC_eqn15}
\end{equation}
Here $u^{\nu}=\mathrm{d}x^{\nu}(t)/\mathrm{d}t$, where the trajectory $x^\mu(t)$ of a fluid element is a geodesic in the Newton–Cartan spacetime.
Taking the spatial components and choosing a comoving reference frame (see Appendix~\ref{A2} for the detailed connection coefficients), the spatial projection of the full NC connection yields
\begin{equation}
\frac{\mathrm{d}u^i}{\mathrm{d}t}+\tilde{\Gamma}_{jk}^iu^ju^k=-\tilde{h}^{ij}\partial_j\phi.
    \label{con:NC_eqn17}
\end{equation}
Substituting $\tilde{h}^{ij} = \mathrm{e}^{-2\phi} \delta^{ij}$ and $\phi = p/\rho$, and using the continuity equation $\tilde{\nabla}_i u^i = 0$ for incompressible flows, the weak-field limit $|\phi| \ll 1$ further reduces equation~\ref{con:NC_eqn17} to
\begin{equation}
\frac{\mathrm{d}u^i}{\mathrm{d}t}=-\frac{1}{\rho}\delta^{ij}\partial_jp,
    \label{con:NC_eqn18}
\end{equation}
which recovers the classical incompressible Euler equation in the weak-field limit. 
This guarantees the self-consistency of the geometric construction with classical fluid mechanics.
It is therefore shown that, in the effective curved space determined by the pressure field, streamlines are geodesics. 
The pressure gradient no longer acts as an external force, and it is absorbed into the geometric effect of the spatial connection.
Unlike Arnold's seminal work \cite{arnold1966geometrie}, which geometrizes the configuration space, the present work geometrizes the physical space.
Since then, we have found this framework well suited for the geometrization of incompressible flow, and therefore the following discussions will focus mainly on the weak limit, i.e., on incompressible flows.

To avoid any confusion about the consequences arising from different geometric effects of pressure on the effective space, it is necessary to distinguish these effects.
The first is the metric scaling effect (see Sect.~\ref{con:NC}), where a high/low-pressure region corresponds to a large/small metric factor, meaning that the effective distance corresponding to a unit coordinate length is larger/smaller. The second is the curvature effect (see Sect.~\ref{con:Rij}), where the effective space curves like a sphere or a saddle surface, causing geodesics to converge or diverge. These two effects are complementary, in that the former describes the local scaling of space (similar to the stretching/compression of a rubber sheet in the general relativity analogy), while the latter describes the way space curves (similar to the curving of a rubber sheet in the general relativity analogy). Together, they determine the behavior of fluid geodesics.

\subsection{Geometrical interpretation of the coupling effect between pressure and velocity}\label{con:Gij}

Applying covariant divergence to both sides of the geodesic equation~\ref{con:NC_eqn17} and realizing that the left hand side of this equation is equivalent to $u^{j}\tilde{\nabla}_{j}u^{i}$ (i.e., $\frac{\mathrm{d}u^i}{\mathrm{d}t}+\tilde{\Gamma}_{jk}^iu^ju^k=\frac{\partial u^i}{\partial t}+u^j\partial_ju^i+\tilde{\Gamma}_{jk}^iu^ju^k=\frac{\partial u^i}{\partial t}+u^{j}\tilde{\nabla}_{j}u^{i}$), the pressure potential Laplacian in the effective space can be obtained:
\begin{equation}
\tilde{\nabla}_{i}u^{j}\tilde{\nabla}_{j}u^{i}+u^{j}\tilde{\nabla}_{i}\tilde{\nabla}_{j}u^{i}=\tilde{\nabla}_{i}(u^{j}\tilde{\nabla}_{j}u^{i})=-\tilde{\nabla}_{i}(\tilde{h}^{ij}\partial_j\phi)=-\tilde{\Box}\phi.
    \label{con:coupling_1}
\end{equation}
Here $\tilde{\Box}=\tilde{h}^{ij}\tilde{\nabla}_i\tilde{\nabla}_j$ is the Laplace operator in curved space.
Again, taking the incompressible continuity equation $\tilde{\nabla}_i u^i = 0$ and the weak-field limit into account, the above equation is reduced to the Poisson equation:
\begin{equation}
\partial_j u^i \partial_i u^j\simeq\tilde{\nabla}_{i}u^{j}\tilde{\nabla}_{j}u^{i}+u^{j}\tilde{\nabla}_{i}\tilde{\nabla}_{j}u^{i}=-\Box\phi \simeq -\frac{1}{\rho}\nabla^2 p.
    \label{con:coupling_2}
\end{equation}
This shows that the classical pressure Poisson equation is not only an independent dynamical derivation but also a derivative of the geodesic equation in the weak-field limit, an inevitable consequence of geometric self-consistency.

\begin{figure*}
	\begin{center}
		\includegraphics[width=0.33\textwidth]{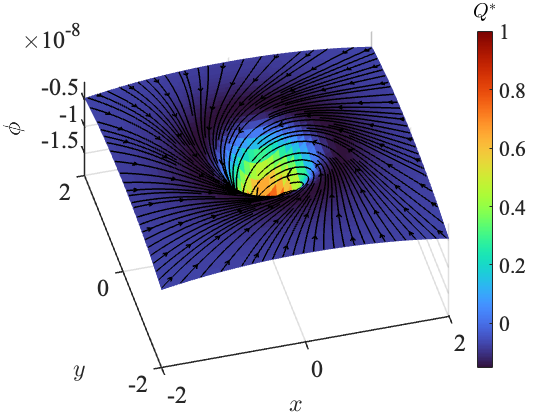}
		\put(-80,-15){(\textbf{a})}	
		\includegraphics[width=0.33\textwidth]{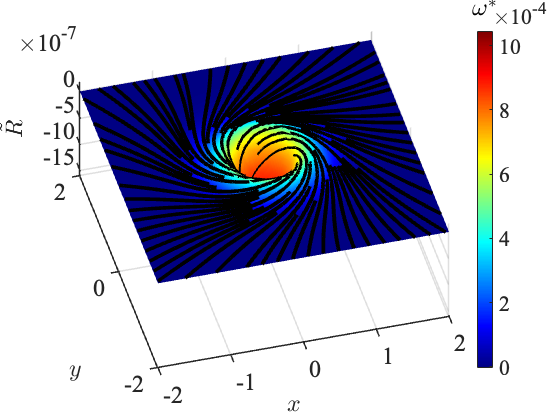}
		\put(-80,-15){(\textbf{b})}
        \includegraphics[width=0.33\textwidth]{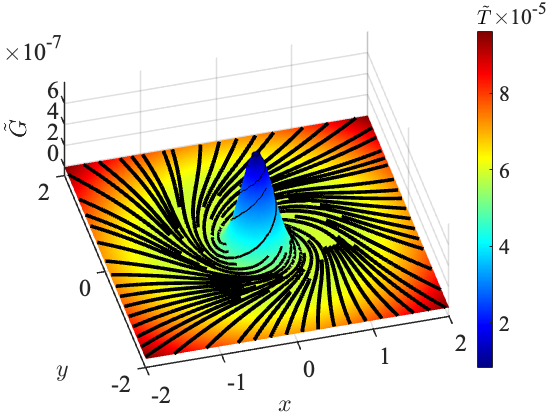}
		\put(-80,-15){(\textbf{c})}
	\end{center}	
	\caption{The Burgers' vortex used in the examination: (a) isosurface of $\phi$ ($z=-0.04$ plane) superimposed with dimensionless $Q^{\ast}=Q/Q_{max}$ contours and streamlines; (b) isosurface of $\tilde{R}$ ($z=-0.04$ plane) superimposed with dimensionless $\omega^{\ast}=\omega l_{c}/c_{0}$ contours and streamlines with linewidth variation according to the metric dilation; (c) isosurface of $\tilde{G}$ ($z=-0.04$ plane) superimposed with vorticity contours and streamlines with linewidth variation according to the metric dilation.}
	\label{Fig:3D_1}
\end{figure*}

The Poisson equation can also be obtained from the perspective of an analogy with the Einstein equations of general relativity, which relate the spatial curvature to the source term (e.g., the stress tensor) to establish a geometric-dynamical analogy.
It is well known that the stress tensor of a fluid is $T_{ij}=\rho u_iu_j+p\delta_{ij}$.
In the current work, the stress tensor $\tilde{T}_{ij}$ in the curved space is expressed as:
\begin{equation}
\tilde{T}_{ij}=\rho u_iu_j+p\tilde{h}_{ij}.
    \label{con:coupling_3}
\end{equation}
Recalling that the Euler equation and geodesics are equivalent in the effective curved space, the application of covariant divergence to $\tilde{T}_{ij}$ gives the steady incompressible Euler equation:
\begin{equation}
\tilde{\nabla}^i\tilde{T}_{ij}=\tilde{\nabla}^i(\rho u_{i}u_{j})+\tilde{\nabla}_{j}p=0.
    \label{con:coupling_4}
\end{equation}
From the basic properties of Riemannian geometry, the spatial tensor satisfies the vanishing covariant divergence $\tilde{\nabla}^{i}\tilde{G}_{ij}=0$ (i.e., the Bianchi identity).
Therefore, the stress tensor $\tilde{T}_{ij}$ and the spatial tensor $\tilde{G}_{ij}$ together should constitute a covariant divergence-free quantity under steady-state conditions:
\begin{equation}
\tilde{\nabla}^i\left(\tilde{G}_{ij}-\lambda \tilde{T}_{ij}\right)=0,
    \label{con:coupling_5}
\end{equation}
where $\lambda$ is a coupling constant.
A more generalized relationship between the spatial curvature and the fluid source can be obtained by taking another spatial covariant divergence $\tilde{\nabla}^j$ of equation~\ref{con:coupling_4} and $\tilde{\nabla}^{i}\tilde{G}_{ij}=0$.
The former again gives the Poisson equation, while the second-order covariant divergences of $\tilde{G}_{ij}$ and $\tilde{T}_{ij}$ together form a field equation:
\begin{equation}
\tilde{\nabla}^i\tilde{\nabla}^j\left(\tilde{G}_{ij}-\kappa \tilde{T}_{ij}\right)=0,
    \label{con:coupling_6}
\end{equation}
where $\kappa$ is a geometric coupling constant. Since the unsteady term vanishes due to the action of $\tilde{\nabla}^j$, equation~\ref{con:coupling_6} is valid for both steady and unsteady conditions. 
This field equation reveals that the structure of the spatial tensor $\tilde{G}_{ij}$ exhibits a striking correspondence with the source terms of the fluid stress tensor. 
It can be seen that pressure appears in both terms of the field equation, which suggests that pressure is not only part of the source generating effective spatial curvature, but is itself a geometric potential describing such curvature.
Therefore, pressure plays a geometric role in fluid flow analogous to that of a ``gravitational charge'' in general relativity.

It is worth noting that the field equation contains no temporal term. This is consistent with the physical nature of pressure in incompressible flows as an instantaneous constraint. In the classical framework, pressure is instantaneously determined by the elliptic Poisson equation to maintain the divergence-free condition of the velocity field. Correspondingly, in the geometric framework, the spatial metric (i.e., the pressure field) is instantaneously determined at each time slice by the field equation, while the dynamical evolution of the system is entirely governed by the geodesic equation. This separation between temporal evolution and instantaneous constraint is precisely the geometric manifestation of the role of pressure in incompressible flows.

This completes the construction of a self-contained theory of pressure geometrization: (i) the fluid source and spatial curvature are correlated via the field equation; (ii) the spatial curvature governs the fluid motion through the geodesic equation; and (iii) the geometric identities automatically ensure momentum conservation and the pressure Poisson equation.

Here we re-examine the 3D Burgers' vortex in the context of the pressure geometrization.
This vortex has been widely used for modeling fine scales of turbulence \cite{pullin_vortex_1998}, whose velocity components in cylindrical coordinates can be written as 
\begin{equation} \label{Burger_expression} 
\begin{split}
  u_{r}& = -\xi r, \\
  u_{\theta}& = \frac{\Gamma}{2\pi r}\left[1-\mathrm{exp}\left(\frac{-r^{2}\xi}{2\nu}\right)\right], \\
  u_{z}& = 2\xi z,
\end{split}
\end{equation}
where $\Gamma$ is the circulation, $\xi$ is the axisymmetric strain rate, and $\nu$ is the kinematic viscosity.
After being transformed into the Cartesian coordinates, the computation domain for the vortex data is $\{x\in\left[-2,2\right], y\in\left[-2,2\right], z\in\left[-2,2\right]\}$ with a characteristic length of $l_{c}=4$. 
Figure~\ref{Fig:3D_1}(a) gives the isosurface of $\phi$ in the $z=-0.04$ plane, superimposed with the dimensionless $Q^{\ast}=Q/Q_{max}$ contours and in-plane streamlines. 
The metric scaling effect caused by the local pressure potential is represented by the dilation/contraction of the streamline linewidth, i.e., the linewidth is linearly mapped from the rescaled (min–max normalization) field of $\phi$.
As expected, the vortex is marked by the concentration of high $Q^{\ast}$ values as well as the low local pressure potential (e.g., the isosurface of $\phi$ is concave).
The low pressure potential is also indicated by the thin streamlines inside the vortex core.
In contrast, thick streamlines outside the vortex core correspond to a large metric factor indicating a relatively high-pressure region.
It is worth noting that the isosurface of $\phi$ is not flat outside the vortex core.
The isosurface of $\tilde{R}$ in the $z=-0.04$ plane superimposed with the dimensionless vorticity $\omega^{\ast}$ is exhibited in figure~\ref{Fig:3D_1}(b).
The vortex is presented as a concave of the $\tilde{R}$ isosurface, indicating that the vortex corresponds to a negative scalar curvature $\tilde{R}$.
Significantly, the $\tilde{R}$ isosurface is purely flat outside the vortex region, where the effective space is intrinsically flat without the curving effect caused by the vortex.
Figure~\ref{Fig:3D_1}(c) exhibits the isosurface of $\tilde{G}$ in the $z=-0.04$ plane superimposed with the trace of the stress tensor $\tilde{T}_{ij}$.
The Burgers' vortex is presented as a convex of the $\tilde{G}$ isosurface, while the $\tilde{G}$ isosurface is, again, purely flat outside the vortex region. 
Combining the distribution of $\tilde{T}$ (i.e., negative in the vortex region) and the field equation~\ref{con:coupling_6}, it can be seen that the spatial tensor $\tilde{G}_{ij}$ and the stress tensor $\tilde{T}_{ij}$ are negatively correlated.
Namely, an increase in $\tilde{G}$ can be interpreted as fluid energy converting into effective spatial curvature, and vice versa.

\begin{figure*}
	\begin{center}
		\includegraphics[width=0.33\textwidth]{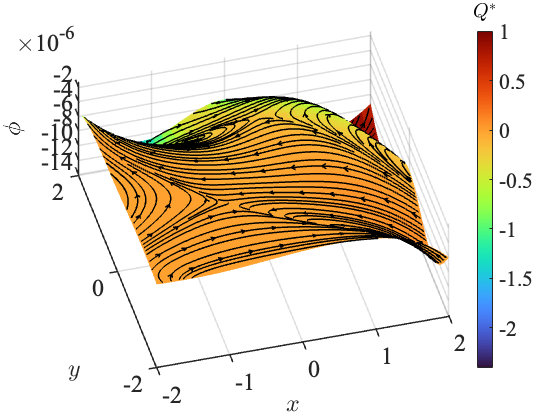}
		\put(-80,-15){(\textbf{a})}	
		\includegraphics[width=0.33\textwidth]{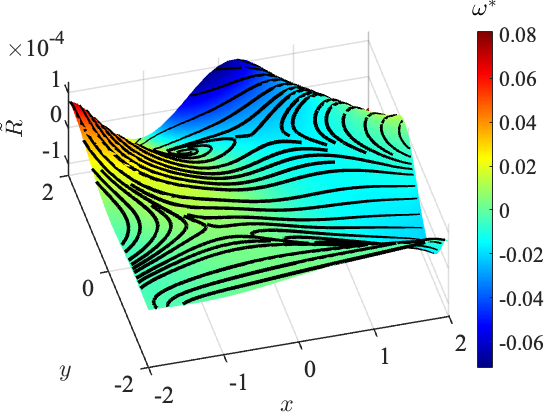}
		\put(-80,-15){(\textbf{b})}
        \includegraphics[width=0.33\textwidth]{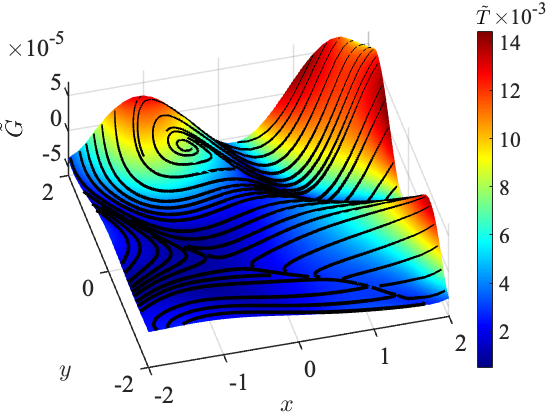}
		\put(-80,-15){(\textbf{c})}
	\end{center}	
	\caption{A Beltrami flow used in the examination: (a) isosurface of $\phi$ ($z=1.875$ plane) superimposed with dimensionless $Q^{\ast}$ contours and streamlines; (b) isosurface of $\tilde{R}$ ($z=1.875$ plane) superimposed with dimensionless $\omega^{\ast}$ contours and streamlines with linewidth variation according to the metric dilation; (c) isosurface of $\tilde{G}$ ($z=1.875$ plane) superimposed with vorticity contours and streamlines with linewidth variation according to the metric dilation.}
	\label{Fig:3D_2}
\end{figure*}

To further demonstrate the idea of the pressure geometrization, we evaluate traces of $\tilde{G}_{ij}$ and $\tilde{T}_{ij}$ on a synthetic dataset generated from an analytical solution proposed by \citet{ethier_exact_1994}. 
This is a three-dimensional unsteady analytical solution to the Navier-Stokes equations, characterized as a Beltrami flow, which is a benchmark commonly used for validating three-dimensional incompressible flow solvers. 
The solution is expressed as
\begin{equation} \label{either_expression} 
\begin{split}
  u& = -a\left[\mathrm{e}^{ax}\mathrm{sin}(ay\pm dz)+\mathrm{e}^{az}\mathrm{cos}(ax\pm dy)\right]\mathrm{e}^{-d^2t}, \\
  v& = -a\left[\mathrm{e}^{ay}\mathrm{sin}(az\pm dx)+\mathrm{e}^{ax}\mathrm{cos}(ay\pm dz)\right]\mathrm{e}^{-d^2t}, \\
  w& = -a\left[\mathrm{e}^{az}\mathrm{sin}(ax\pm dy)+\mathrm{e}^{ay}\mathrm{cos}(az\pm dx)\right]\mathrm{e}^{-d^2t},
\end{split}
\end{equation}
where $a^2+b^2=0$. A family of velocity fields of a Beltrami flow can be obtained depending on the selection of $a$ and $b$.
The computation domain is $\{x\in\left[-2,2\right], y\in\left[-2,2\right], z\in\left[-2,2\right]\}$.
Similar to figure~\ref{Fig:3D_1}, figure~\ref{Fig:3D_2} exhibits the isosurfaces of $\phi$, $\tilde{R}$ and $\tilde{G}$ in the $z=1.875$ plane, superimposed with contours of $Q^{\ast}$, vorticity $\omega^{\ast}$ and $\tilde{T}$, respectively.

A saddle point is located at $\{x=-0.75,y=-0.875\}$, indicated by the converging streamlines in figure~\ref{Fig:3D_2}(a). 
Still, the pressure potential isosurface exhibits a relatively ``smooth'' distribution, i.e., no obvious characteristic structure in the potential level to reveal the local convergence of the flow.
In contrast, asymptotes of the converging streamlines are presented as a ridge and a concave for the $\tilde{R}$ and $\tilde{G}$ isosurfaces, respectively. 
Therefore, shear effects are also resultants of the effective spatial curvature, though they have an opposite curving direction to that of vortices.
Around $\{x=-0.73,y=1.22\}$, a vortical structure is indicated by local spiral streamlines.
Again, in terms of the isosurface, the vortex is presented as a concave of $\phi$ or $\tilde{R}$, or a convex of $\tilde{G}$.
Yet interestingly, the spiral node indicated by the streamlines is neither located on an apex of the isosurface of $\tilde{G}$, nor at the local minimum of the contour of $\tilde{T}$.
Such a peak-shift between the pressure and velocity fields \cite{liang2026investigation} is implied by the difference between the stress tensor and the spatial tensor, which is a tensor with zero second-order covariant divergence (see equation~\ref{con:coupling_6}).

\section{Degeneration to two-dimensional flows}\label{con:2D}

An important scenario is the degeneration to two-dimensional flows, where the Ricci tensor and the effective spatial tensor undergo significant variations since $\delta^i_i=2$ rather than 3.
Consequently, it is necessary to investigate the geometrization characteristics of the degenerate flow.

For a two-dimensional conformally flat metric, the Ricci tensor gives
\begin{equation}
\tilde{R}^\mathrm{2D}_{ij}=-\delta_{ij}\nabla^2\phi,
    \label{con:NC_eqn19}
\end{equation}
and the corresponding trace is given as
\begin{equation}
\tilde{R}^\mathrm{2D}=\tilde{h}^{ij}\tilde{R}^\mathrm{2D}_{ij}=-2\mathrm{e}^{-2\phi}\nabla^2\phi.
    \label{con:NC_eqn20}
\end{equation}
It can be seen that, for 2D flows, the spatial curvature is exactly represented by the pressure Laplacian, and the dependence of the pressure gradient on orientation (i.e., the pressure Hessian) is completely canceled by the nonlinear terms of the connection.

It is well known that the pressure Laplacian can be further decomposed into the symmetric rate of strain tensor $\boldsymbol{S}$ and the antisymmetric rate of rotation tensor $\boldsymbol{\Omega}$, as $\nabla^2p=-\rho(S^{ij}S_{ij}-\Omega^{ij}\Omega_{ij})=-\rho(S^{ij}S_{ij}-\frac{1}{2}|\boldsymbol{\omega}|^2)$.
Thus, the Ricci tensor in the 2D case directly determines the dynamic behavior of the flow, whether it is rotation-dominated or stretching-dominated. 
Specifically, a positive curvature (i.e., $\tilde{R}>0$) indicates strong strain effect (geodesics converging), while a negative curvature (i.e., $\tilde{R}<0$) indicates strong rotating effect (geodesics paralleling/diverging). 
In addition, the absence of the pressure Hessian suggests that the effective spatial curvature caused by pressure is isotropic (uniquely determined by the scalar $\nabla^2p$), whereas the curvature of three-dimensional flow is anisotropic (the direction is determined by the tensor $\partial_i\partial_jp$). 
Such phenomenon is consistent with the fact that 3D flow exhibits more complex vortex structures, such as vortex stretching and tilting.

More importantly, the effective spatial tensor $\tilde{G}^\mathrm{2D}_{ij}$ vanishes in 2D flows, reflecting the degeneracy inherent in the geometric structure:
\begin{equation}
\tilde{G}^\mathrm{2D}_{ij}=\tilde{R}^\mathrm{2D}_{ij}-\frac{1}{2}\tilde{R}^\mathrm{2D}\tilde{h}_{ij}=0.
    \label{con:NC_eqn21}
\end{equation}
This indicates that the curvature of the 2D effective space is strictly restricted, allowing only for directionally uniform stretching or contraction.
Such rigidity constrains the direction of energy transfer, so that positive-curvature condensation (geodesics converging due to the positive curvature $\tilde{R}_{+}$) tends to merge toward larger scales similar to how gravity leads to the clustering of matter in general relativity, while negative-curvature dispersion (geodesics diverging due to the negative curvature $\tilde{R}_{-}$) tends to fragment toward smaller scales. 
Furthermore, by denoting enstrophy as $\mathscr{E}=\frac{1}{2}\int |\boldsymbol{\omega}|^2 \mathrm{d}^2x$ and recalling the identity in the 2D condition $\int |\nabla \boldsymbol{u}|^2 \mathrm{d}^2x=\int \left(\mathrm{tr}(\boldsymbol{S}^2)+\frac{1}{2}|\boldsymbol{\omega}|^2\right)\mathrm{d}^2x=2\mathscr{E}$, we can see that the integral of the total effective curvature vanishes in a 2D inviscid flow:
\begin{equation}
\int\tilde{R}^{\mathrm{2D}}\mathrm{d}^2x \varpropto \int\left(\mathrm{tr}(\boldsymbol{S}^2)-\frac{1}{2}|\boldsymbol{\omega}|^2\right)\mathrm{d}^2x=\mathscr{E}-\mathscr{E}=0.
    \label{con:NC_eqn22}
\end{equation}
In other words, though the spatial distributions of $\tilde{R}_{+}$ and $\tilde{R}_{-}$ are different, the net curvature must balance globally (for an infinite domain or periodic boundaries).

It is well known that the reduction in dimensionality leads to the dual-cascade scenario with an inverse energy cascade to larger scales and a direct enstrophy cascade to smaller scales for 2D turbulent flows \cite{kraichnan_inertial_1967,annurev2Dtur}.
In the scope of geometrization proposed by the current work, the inverse energy cascade is not the conversion of energy itself into positive curvature, but rather the geometric condensation of positive-curvature structures toward larger scales, which leads to the growth of energy at large scales. Likewise, the forward enstrophy cascade is not the conversion of vorticity itself into negative curvature, but rather the geometric dispersion caused by negative-curvature structures, which allows vorticity to be transferred toward smaller scales.
The underlying reason for the coexistence of the inverse energy cascade and the forward enstrophy cascade may lie exactly in the global topological constraint on curvature (see equation~\ref{con:NC_eqn22}), namely that $\tilde{R}_{+}$ and $\tilde{R}_{-}$ cannot evolve separately, and that their integrals must always be kept in equilibrium.

\begin{figure*}
	\begin{center}
		\includegraphics[width=0.33\textwidth]{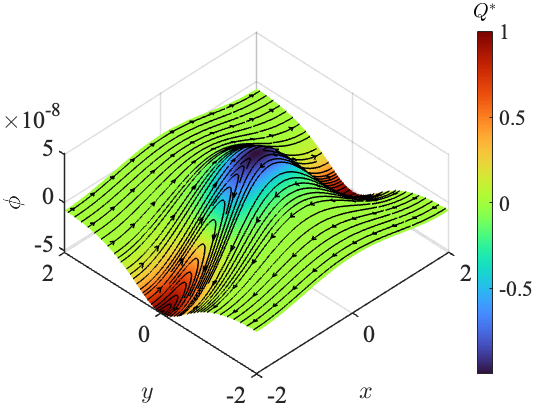}
		\put(-80,-15){(\textbf{a})}	
		\includegraphics[width=0.33\textwidth]{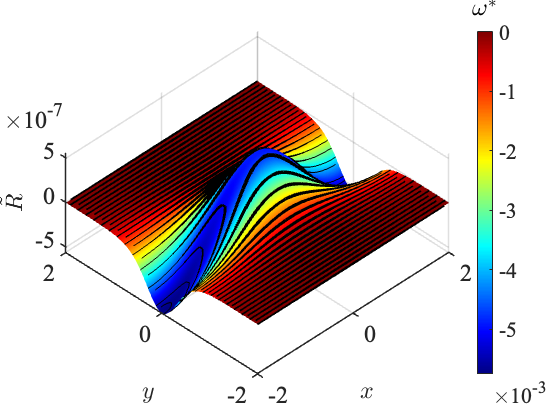}
		\put(-80,-15){(\textbf{b})}
        \includegraphics[width=0.33\textwidth]{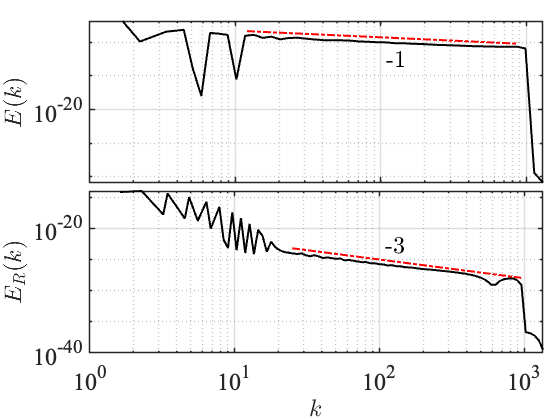}
		\put(-80,-15){(\textbf{c})}
	\end{center}	
	\caption{A shear flow used in the 2D examination: (a) isosurface of $\phi$ superimposed with dimensionless $Q^{\ast}$ contours and streamlines; (b) isosurface of $\tilde{R}$ superimposed with dimensionless $\omega^{\ast}$ contours and streamlines with linewidth variation according to the metric dilation; (c) energy and curvature spectra.}
	\label{Fig:degenerate_to_2D_1}
\end{figure*}

To exhibit the morphological characteristics of the effective spatial curvature in the degenerate scenario, we re-examine two typical 2D ideal flows, i.e., a 2D shear flow and a 2D dual-vortex flow.
The flowfield of the former is expressed as
\begin{equation} \label{shear_expression} 
\begin{split}
  u&=\mathrm{tanh}(2y), \\
  v& = 0.1\mathrm{sin}(2\pi x)\mathrm{e}^{-y^2}.
\end{split}
\end{equation}
And the flowfield of the latter is written as
\begin{equation} \label{dualvortex_expression} 
\begin{split}
  u&=\frac{\Gamma_{1}(y-y_{1})}{2\pi\sigma^2}\mathrm{e}^{-\left[\frac{(x-x_1)^2+(y-y_1)^2}{2\sigma^2}\right]}+\frac{\Gamma_{2}(y-y_{2})}{2\pi\sigma^2}\mathrm{e}^{-\left[\frac{(x-x_2)^2+(y-y_2)^2}{2\sigma^2}\right]},
  \\
  v&=-\frac{\Gamma_{1}(y-y_{1})}{2\pi\sigma^2}\mathrm{e}^{-\left[\frac{(x-x_1)^2+(y-y_1)^2}{2\sigma^2}\right]}-\frac{\Gamma_{2}(y-y_{2})}{2\pi\sigma^2}\mathrm{e}^{-\left[\frac{(x-x_2)^2+(y-y_2)^2}{2\sigma^2}\right]}.
\end{split}
\end{equation}
Here $(x_1,y_1)$ and $(x_2,y_2)$ denote the coordinates of the centers of the vortices, $\Gamma_1$ and $\Gamma_2$ describe the circulations of the vortices, and $\sigma$ is the vortex radius.
The computation domain for both cases is $\{x\in\left[-2,2\right],y\in\left[-2,2\right]\}$.

\begin{figure*}
	\begin{center}       
        \includegraphics[width=0.33\textwidth]{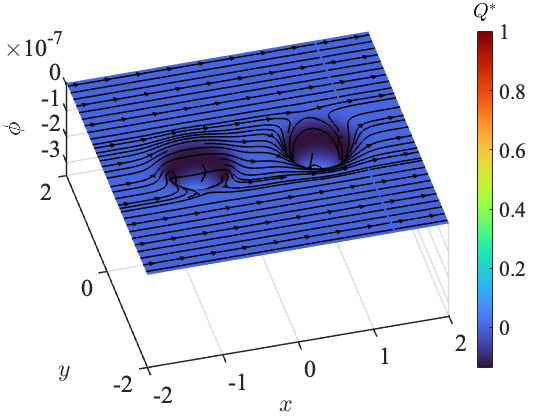}
		\put(-80,-15){(\textbf{a})}
        \includegraphics[width=0.33\textwidth]{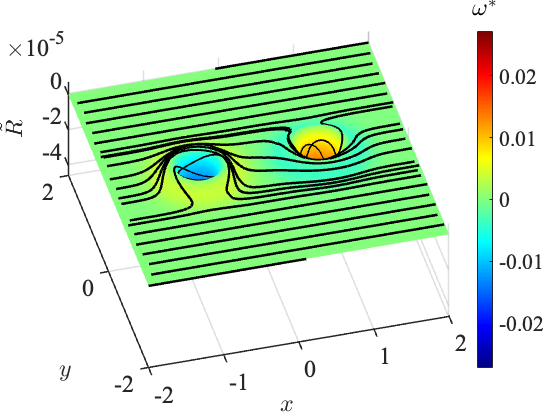}
		\put(-80,-15){(\textbf{b})}
        \includegraphics[width=0.33\textwidth]{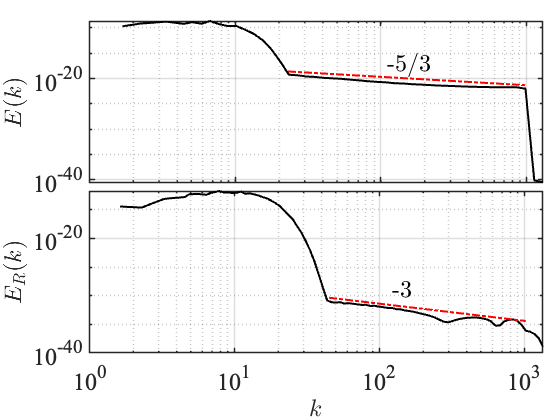}
		\put(-80,-15){(\textbf{c})}
	\end{center}	
	\caption{A dual-vortex flow used in the 2D examination: (a) isosurface of $\phi$ superimposed with dimensionless $Q^{\ast}$ contours and streamlines; (b) isosurface of $\tilde{R}$ superimposed with dimensionless $\omega^{\ast}$ contours and streamlines with linewidth variation according to the metric dilation; (c) energy and curvature spectra.}
	\label{Fig:degenerate_to_2D_2}
\end{figure*}

Following the paradigm of figures~\ref{Fig:3D_1} and \ref{Fig:3D_2}, figures~\ref{Fig:degenerate_to_2D_1} and \ref{Fig:degenerate_to_2D_2} give the comparison of the pressure potential and curvature fields.
As shown in figure~\ref{Fig:degenerate_to_2D_1}(a), two vortices and a saddle are formed between two layers of high speed flows. These vortices are distinguished by the positive $Q^{\ast}$ and the pressure potential trough, while the saddle is indicated by the negative $Q^{\ast}$ and the pressure potential plateau. 
Again, the isosurface of $\tilde{R}$ (figure~\ref{Fig:degenerate_to_2D_1}b) is purely flat outside the regions of these coherent structures.
The coherent structures can be regarded as intrinsic curvatures of the effective space.
Figure~\ref{Fig:degenerate_to_2D_1}(c) gives the energy spectra of the energy cascade $E(k)$ and the curvature cascade $E_{R}(k)$ based on $\tilde{R}$.
Since the examined shear flow is not a standard 2D turbulent flow, its energy cascade behaves with an order of $k^{-1}$.
The curvature cascade $E_{R}(k)\sim k^{-3}$ implies that the curvature energy spectrum is predominantly concentrated at large scales (i.e., at small $k$).
Figure~\ref{Fig:degenerate_to_2D_2} exhibits the isosurfaces of $\phi$ and $\tilde{R}$ of the dual-vortex flow. As expected, the vortices are presented as concavities of the $\tilde{R}$ isosurface, indicating that a vortex corresponds to a negative scalar curvature $\tilde{R}$ causing geometric dispersion and transformation of vorticity toward smaller scales.
With a standard inverse energy cascade $E(k)\sim k^{-5/3}$, the curvature cascade also gives the $k^{-3}$ spectrum (figure~\ref{Fig:degenerate_to_2D_2}c).

\begin{figure}
	\begin{center}       
        \includegraphics[width=0.48\textwidth]{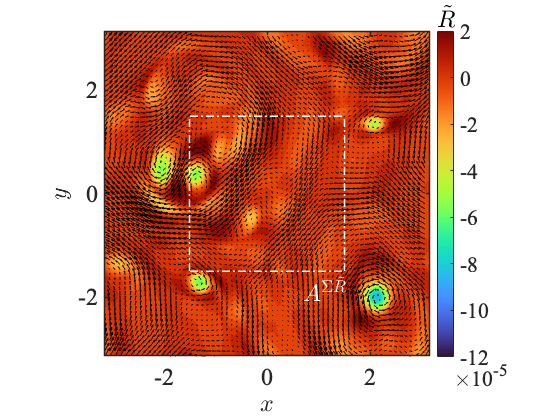}
		\put(-120,-10){(\textbf{a})}\\
        \includegraphics[width=0.4\textwidth]{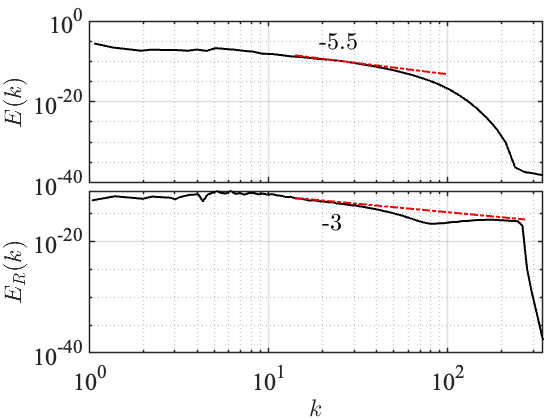}
		\put(-100,-10){(\textbf{b})}
	\end{center}	
	\caption{DNS of 2D turbulence used in the 2D examination: (a) $\tilde{R}$ contours superimposed with velocity vectors, with an inset box indicating the integral area $A^{\Sigma\tilde{R}}$ of the net curvature; (b) energy and curvature spectra.}
	\label{Fig:degenerate_to_2D_3}
\end{figure}

To further examine the characteristics of the effective spatial curvature in 2D turbulence, the distribution of $\tilde{R}$ and the cascades of energy and curvature are studied and presented in figure~\ref{Fig:degenerate_to_2D_3}.
The data are generated from direct numerical simulation (DNS) of 2D turbulent forced fluid flow provided in \citet{Vladimir2024}.
The corresponding data sets can be downloaded from the website: \url{https://github.com/parfenyev/2d-turb-PINN/tree/main/PINN}.
Apparently, vortices correspond to concavities of $\tilde{R}$ (figure~\ref{Fig:degenerate_to_2D_3}a).
While the energy cascade of the forced 2D turbulence behaves with an order of $k^{-5.5}$, the curvature cascade again presents the $k^{-3}$ spectrum (figure~\ref{Fig:degenerate_to_2D_3}b).
Such a spectrum indicates that the fluid condenses at large scales due to the effective curvature, which may lead to the inverse cascade of energy.

\begin{figure*}
	\begin{center}       
        \includegraphics[width=0.33\textwidth]{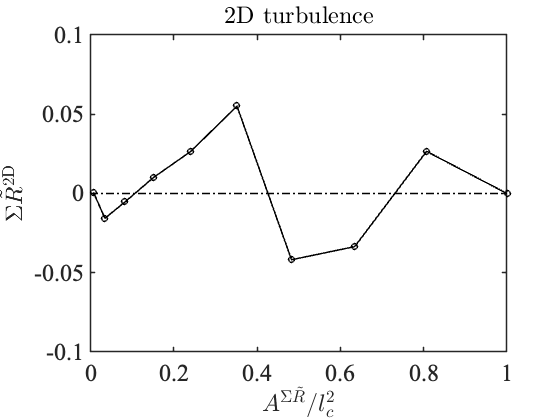}
		\put(-80,-15){(\textbf{a})}
        \includegraphics[width=0.33\textwidth]{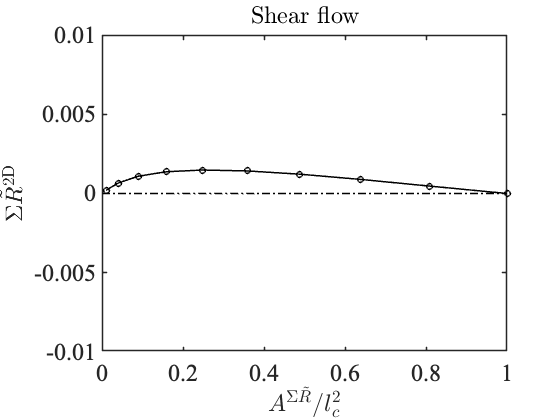}
		\put(-80,-15){(\textbf{b})}
        \includegraphics[width=0.33\textwidth]{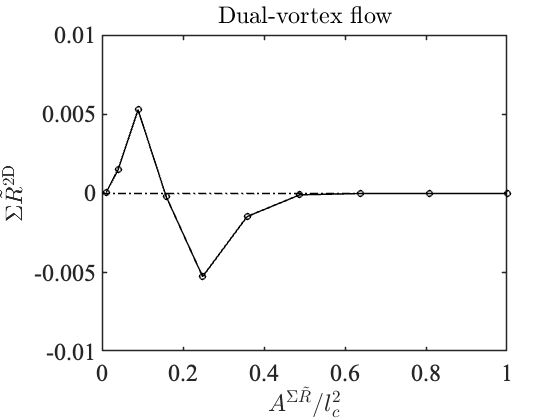}
		\put(-80,-15){(\textbf{c})}
	\end{center}	
	\caption{Variation in the net curvature $\Sigma\tilde{R}^{\mathrm{2D}}$ with increasing integral area $A^{\Sigma\tilde{R}}$: (a) 2D turbulence; (b) shear flow; (c) dual-vortex flow.}
	\label{Fig:degenerate_to_2D_4}
\end{figure*}

Figure~\ref{Fig:degenerate_to_2D_4} shows how the net curvature varies with increasing integral area $A^{\Sigma\tilde{R}}$ for the shear, dual-vortex, and 2D turbulent flows.
As indicated by the inset box in figure~\ref{Fig:degenerate_to_2D_3}(a), $A^{\Sigma\tilde{R}}$ starts at the center of the computation domain and grows until it covers the entire domain.
Thus, $A^{\Sigma\tilde{R}}/l^2_{c}$ represents the ratio of the integral area to the computation domain.
Since all $\Sigma\tilde{R}^{\mathrm{2D}}$ curves approach zero, $\tilde{R}_{+}$ and $\tilde{R}_{-}$ tend to cancel each other out as the integral area increases.
In other words, the net curvature tends to be globally balanced.

Combining equation~\ref{con:NC_eqn22} and the above observation, it can be concluded that large-scale spatial curvature dominates the geometric structure of the flow, and the net curvature cancels out after global integration of multi-scale curvatures.
Consequently, the inverse cascade due to geometric condensation of $\tilde{R}_{+}$ and the forward cascade due to geometric dispersion of $\tilde{R}_{-}$ coexist but at different scales.  

\section{Geometric interpretation of prevailing vortex identification schemes}\label{pre_vid}

Although the identification of vortices is not a new problem, a widely accepted, objective definition 
of a vortex remains elusive \cite{Jeong_Hussain_1995}.
Prevailing vortex identification schemes are generally based on closed or spiraling streamlines and pathlines, local pressure minima, and VGT decomposition from the Eulerian perspective \cite{chakraborty_relationships_2005}, or flow maps from the Lagrangian perspective \cite{haller_lagrangian_2002,haller_lagrangian_2015}.
In this section, we provide a geometric interpretation of these schemes and show that they actually correspond to different aspects of the effective spatial curvature.

\subsection{Q criterion}\label{Q}
The $Q$ criterion identifies regions as vortices if the norm of the local rate of rotation tensor is dominant over the norm of the local rate of strain tensor \cite{moin_eddies_1988}. 
As a typical vortex identification method based on the VGT, it is defined as,
\begin{equation}
Q=\frac{1}{2}\left(||\boldsymbol{\Omega}||^2-||\boldsymbol{S}||^2\right),
    \label{Q_1}
\end{equation}
where $||\cdot||$ denotes the Frobenius norm.
A vortex is defined in those regions where $Q>0$, which is interpreted as a dominance
of rotation over strain. For incompressible flows, $Q$ is defined as the second invariant of the VGT, which again relates to the pressure Laplacian:
\begin{equation}
Q=\frac{1}{2}\left[(\partial^iu_i)^2-\partial^iu_j\partial^ju_i\right]=-\frac{1}{2}\partial^iu_j\partial^ju_i=\frac{1}{2\rho}\nabla^2p.
    \label{Q_2}
\end{equation}

For 3D incompressible flows, the effective spatial curvature gives $\tilde{G}\simeq2\nabla^2\phi=\frac{2}{\rho}\nabla^2p$. Thus, we can obtain
\begin{equation}
Q=\frac{1}{4}\tilde{G}.
    \label{Q_3}
\end{equation}
Therefore, the $Q$ criterion actually quantifies the trace of the effective spatial tensor $\tilde{G}$.
Since $\tilde{G}$ is a geometric quantity directly determined by the matter source (see equation~\ref{con:coupling_6}), the existence of a vortex ($Q>0$, dominated by the rotary effect) requires a positive spatial curvature-energy coupling ($\tilde{G}>0$, a positive effective spatial curvature determined by the stress-energy-momentum tensor). 
Conversely, the domination of strain ($Q<0$) corresponds to a negative coupling ($\tilde{G}<0$).

For degenerate 2D flows, the effective spatial curvature vanishes while the Ricci tensor takes over the place of curvature determination ($\tilde{R}^{\mathrm{2D}}\simeq2\nabla^2\phi$, see equation~\ref{con:NC_eqn20}). 
Interestingly, the $Q$ criterion is proportional to the negative value of the intrinsic spatial curvature:
\begin{equation}
Q=-\frac{1}{4}\tilde{R}^{\mathrm{2D}}.
    \label{Q_4}
\end{equation}
In short, the $Q$ criterion exhibits the coupling effect for 3D flows, and it exhibits the intrinsic spatial curving effect for 2D flows.

Additionally, based on equations~\ref{Q_1}, \ref{Q_3} and the relevant discussion, vorticity can be regarded as the source generating positive/negative curvature of $\tilde{G}_{ij}$/$\tilde{T}_{ij}$ by noting that $|\boldsymbol{\omega}|^2-2\mathrm{tr}(\boldsymbol{S}^2)=\tilde{G}=-\tilde{R}/2$.

\subsection{$\lambda_{ci}$ criterion}\label{lambda2}

The $\lambda_{ci}$ criterion \cite{zhou_mechanisms_1999} is discussed after the $Q$ criterion due to their high similarity in geometric interpretation.
Such a swirling strength criterion uses the imaginary part of the complex conjugate eigenvalue ($\lambda_{cr}\pm\lambda_{ci}i$) of the VGT to identify vortices.
Specifically, when $\nabla\boldsymbol{u}$ has complex conjugate eigenvalues, in the locally curvilinear coordinate system $\left(y_1, y_2, y_3\right)$ spanned by the vectors $\left(\boldsymbol{\nu}_{r},\boldsymbol{\nu}_{cr},\boldsymbol{\nu}_{ci}\right)$, the flow should be locally swirling in the plane spanned by $\left(\boldsymbol{\nu}_{cr},\boldsymbol{\nu}_{ci}\right)$ and is stretched or compressed along $\boldsymbol{\nu}_{r}$. 

For 3D flows, the characteristic equation for $\nabla\boldsymbol{u}$ is given by
\begin{equation}
\lambda^{3}+P\lambda^{2}+Q\lambda+R=0,
    \label{ci_1}
\end{equation}
where $P$, $Q$ and $R$ are the three invariants of $\nabla\boldsymbol{u}$, defined as $P=-\nabla\cdot\boldsymbol{u}$ (first invariant), $Q$ (second invariant defined in Sect.~\ref{Q}), and $R=-\mathrm{Det}(\nabla\boldsymbol{u})$ (third invariant). 
The discriminant for equation~\ref{ci_1} is (for incompressible flows, i.e., when $P=0$)
\begin{equation}
\Delta=\left(\frac{1}{2}R\right)^{2}+\left(\frac{1}{3}Q\right)^{3}.
    \label{ci_2}
\end{equation}
The conditions $\Delta>0$ and $Q>0$ imply that $\nabla\boldsymbol{u}$ has complex eigenvalues. 
Then $\lambda_{ci}$ can be solved by recalling that $Q=\tilde{G}/4$ (equation~\ref{Q_3}):
\begin{equation}
\lambda_{ci}=\frac{1}{2}\sqrt{\tilde{G}}\mathrm{cosh}\left(\frac{\theta}{3}\right),
    \label{ci_3}
\end{equation}
where $\theta=\mathrm{sinh}^{-1}\left(\frac{-4R}{(\tilde{G}/3)^{3/2}}\right)$.
Namely, the swirling strength is proportional to the square root of the effective spatial curvature with a modification factor $\mathrm{cosh}(\theta /3)$, which is a coupling quantity of curvature with the third invariant, characterizing the extent of vortex stretching.

For degenerate 2D flows, the characteristic equation for $\nabla\boldsymbol{u}$ reduces to
\begin{equation}
\lambda^{2}+Q=0.
    \label{ci_4}
\end{equation}
With the discriminant simplified to $\Delta=-4Q$, $\lambda_{ci}$ can be directly obtained by recalling that $Q=-\tilde{R}^{\mathrm{2D}}/4$ for 2D flows (equation~\ref{Q_4}):
\begin{equation}
\lambda_{ci}=\frac{1}{2}\sqrt{-\tilde{R}^{\mathrm{2D}}}.
    \label{ci_5}
\end{equation}

Equations~\ref{ci_3} and \ref{ci_5} again exhibit the fundamental difference between 3D and 2D vortices, namely that no stretching exists in 2D vortices.
When the stretching effect is much weaker than the curving effect (i.e., $|R/16|\ll|(\tilde{G}/3)^{3/2}|$), the modification factor approaches one and the 3D vortex approximates a 2D vortex:
\begin{equation}
\lambda_{ci}\mathrel{\big|}_{\theta\rightarrow0}=\frac{1}{2}\sqrt{\tilde{G}}=\frac{1}{2}\sqrt{-\tilde{R}^{\mathrm{2D}}},
    \label{ci_6}
\end{equation}
by noting that the Ricci tensor takes over the place of curvature determination in the 2D scenario. 

In short, the $\lambda_{ci}$ criterion corresponds to the specific mode of the affine connection exhibiting the coupling/curving effect, which quantifies the degree of the effective spatial curvature.

\subsection{$\lambda_{2}$ criterion}\label{lambdaci}

The $\lambda_{2}$ criterion \cite{Jeong_Hussain_1995} is formulated based on the observation that the local pressure minimum in a plane is perpendicular to the vortex axis by neglecting unsteady and viscous effects.
The symmetric part of the gradient of the incompressible Navier-Stokes equation can
be expressed as
\begin{equation}
\boldsymbol{S}^2+\boldsymbol{\Omega}^2=-\frac{1}{\rho}\partial^{2}_{ij}p.
    \label{c2_1}
\end{equation}
The vortex core is defined as a connected region with two positive eigenvalues of the pressure Hessian $\partial^{2}_{ij}p$.
If the eigenvalues of the pressure Hessian (or the symmetric tensor $\boldsymbol{S}^2+\boldsymbol{\Omega}^2$) are ordered as $\lambda_{1}\ge\lambda_{2}\ge\lambda_{3}$, then the vortex is defined as regions satisfying the requirement that $\lambda_{2}<0$.

It has been shown that the effective spatial tensor $\tilde{G}_{ij}$ is made up of the pressure Hessian, the pressure Laplacian and a higher order term (equation~\ref{con:NC_eqn9}).
The symmetric tensor $\boldsymbol{S}^2+\boldsymbol{\Omega}^2$ therefore can be converted into a linear combination of the spatial/Ricci tensor and the scalar curvature (see Appendix \ref{A3.5}):
\begin{equation}
\boldsymbol{S}^2+\boldsymbol{\Omega}^2=\left(\tilde{G}_{ij}-\frac{1}{2}\tilde{G}\tilde{h}_{ij}\right)=\left(\tilde{R}_{ij}-\frac{1}{4}\tilde{R}\tilde{h}_{ij}\right).
    \label{c2_2}
\end{equation}

In the geometric language, the $\lambda_2$ criterion identifies regions where the intrinsic spatial curvature, when projected onto appropriate planes (e.g., $\tilde{G}_{ij}-\frac{1}{2}\tilde{G}\tilde{h}_{ij}$ or $\tilde{R}_{ij}-\frac{1}{4}\tilde{R}\tilde{h}_{ij}$), is negative. 
The $\lambda_{2}$ criterion thus demarcates the region where the curved space in the projected plane becomes hyperbolic enough to sustain the coherent swirling motion.
However, different from $\tilde{G}_{ij}$ or $\tilde{R}_{ij}$, the $\lambda_{2}$ criterion does not take the pressure Laplacian into account, which may lose information about the total scalar curvature induced by the pressure field.

\subsection{Finite-time Lyapunov exponent}\label{LCS}

Lagrangian coherent structure (LCS) analysis was described by \citet{HALLER2001248}, and includes a series of Lagrangian methods that calculate quantities based on the behavior of fluid particle trajectories \cite{huang_detection_2015}.
One such method identifies LCS as maximizing ridges of the finite-time Lyapunov exponent (FTLE) field.
It is a scalar quantity calculated at each point in space ($\boldsymbol{x}_{0}$) that measures maximum separation rate among neighboring particle trajectories initialized near that point.
The flow map $\Phi(\boldsymbol{x}_{0},t_{0},T)$ is first calculated as the vector location of the corresponding trajectory initiated at $(\boldsymbol{x}_{0},t_{0})$ at its final location over an integration time $T$.
The Cauchy-Green strain tensor $C(\boldsymbol{x}_{0})$ is then constructed from the spatial gradient of the flow map.
The maximum eigenvalue of the Cauchy–Green strain tensor is referred to as the coefficient of expansion $\sigma_{T}$,
\begin{equation}
\begin{split}
\sigma_{T}(\boldsymbol{x}_{0},t_{0})&=\lambda_{max}\left(C(\boldsymbol{x}_{0})\right)\\&=\lambda_{max}\Bigg (\bigg [\frac{\partial \Phi(\boldsymbol{x}_{0},t_{0},T)}{\partial \boldsymbol{x}_{0}}\bigg ]^{T}\bigg [\frac{\partial \Phi(\boldsymbol{x}_{0},t_{0},T)}{\partial \boldsymbol{x}_{0}}\bigg ]\Bigg).
 \label{ftle1}
\end{split}
\end{equation}
The FTLE field is then defined as,
\begin{equation}
\text{FTLE}_{T}(\boldsymbol{x}_{0},t_{0})=\frac{1}{T}\log \sqrt{\sigma_{T}(\boldsymbol{x}_{0},t_{0})}.
 \label{ftle2}
\end{equation}
Maximizing ridges in the scalar FTLE field represent strong material stretching among nearby particle trajectories \cite{green2011,tu_ftle_2022}.

The attracting/repelling of trajectories of neighboring particles measured by FTLE is equivalent to the convergence/divergence of geodesics in the effective space, which is prescribed by the geodesic deviation equation
\begin{equation}
\frac{\mathrm{d}^2 \xi^{i}}{\mathrm{d}t^2}+\tilde{R}^{i}_{jkl}u^{j}\xi^{k}u^{l}=0.
    \label{lcs_1}
\end{equation}
Here $\xi^{i}$ is the separation vector indicating the separation of neighboring geodesics, and $\tilde{R}^{i}_{jkl}$ is the Riemann curvature tensor determined by the Ricci tensor (see Appendix~\ref{A4}):
\begin{equation}
\tilde{R}_{ijkl}=-\delta_{il}\partial_{j}\partial_{k}\phi+\delta_{ik}\partial_{j}\partial_{l}\phi-\delta_{jk}\partial_{i}\partial_{l}\phi+\delta_{jl}\partial_{i}\partial_{k}\phi.
    \label{lcs_2}
\end{equation}
By noting the linear relationship between the pressure Hessian and the terms in equation~\ref{lcs_2} (i.e., $\partial^{2}_{ij}p=\rho\partial_{i}\partial_{j}\phi$), it can be seen that the rate of separation of neighboring trajectories should be governed by the projection of the pressure Hessian onto the geodesic:
\begin{equation}
\sigma \varpropto \sqrt{|\tilde{R}^{i}_{jkl}u^{j}u^{l}|}.
    \label{lcs_3}
\end{equation}
Therefore, the ridges of the FTLE correspond to regions of maximum eigenvalues of the tensor $\tilde{R}^{i}_{jkl}u^{j}u^{l}$.
The behavior of fluid particle trajectories can be attributed to geodesic deviation forced by the effective spatial curvature induced by pressure. 
Building upon this picture, strong separation may arise from a parallelism effect. When the local velocity vector aligns with an eigenvector of the Riemann tensor, the particle closely shadows a particular geodesic, so that its deviation from that geodesic is reduced, whereas neighboring particles do not. 
As a result, the relative separation between them is enhanced. 
Such an interpretive conjecture is consistent with the observation of \citet{haller2012geodesic} that the locally most compressing or most shearing material lines are closely shadowed by least-stretching geodesics of the metric induced by the Cauchy-Green strain tensor.





\begin{figure*}
	\begin{center}       
        \includegraphics[width=1\textwidth]{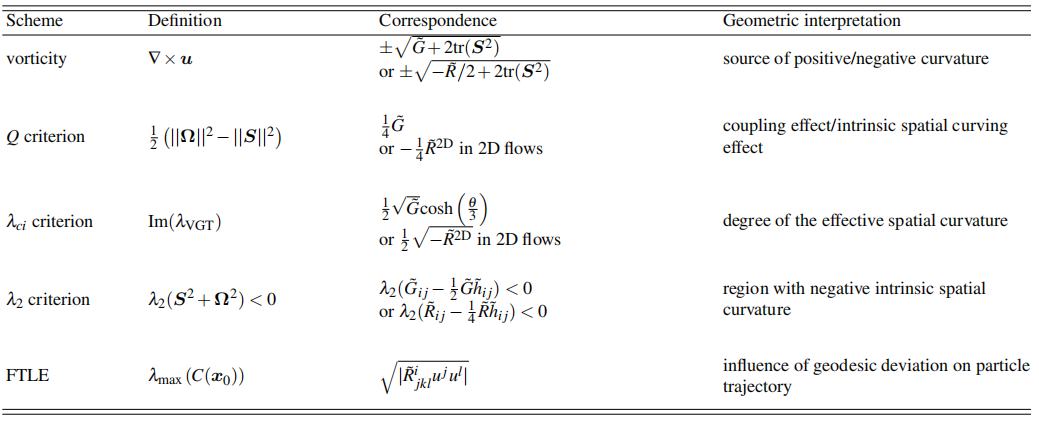}
	\end{center}	
	\caption{An overview of geometric interpretation of current vortex identification schemes.}
	\label{Tab:lcs}
\end{figure*}

To end this section, we summarize the interpretations of the aforementioned schemes in table~\ref{Tab:lcs} to give an overall geometric scope of prevailing vortex identification approaches.

\section{Discussions}\label{Discussions}


In this paper, we propose a novel theoretical framework for geometrizing pressure in an analogous way to general relativity theory.
By establishing a Newton-Cartan geometry and applying a conformal transformation, we show that the pressure field can be consistently interpreted as a geometric potential that induces an effective spatial curvature.
This curvature, encoded in tensors such as the Ricci tensor $\tilde{R}_{ij}$ and the Einstein-like spatial tensor $\tilde{G}_{ij}$, is not merely a mathematical abstraction, but also directly governs fluid motion through the geodesic equation.
The curvature is dynamically coupled to the source terms, mirroring the profound structure of the Einstein field equations.
Several significant insights can be drawn from the current geometrization framework, which enrich our understanding of pressure and vortex dynamics.


A pivotal implication of this geometrization is the redefinition of the physical role of pressure.
In the classical paradigm, pressure appears as a scalar field enforcing the incompressibility constraint via the Poisson equation, and its gradient manifests as an external force in the momentum balance.
Within the present geometric framework, however, pressure is entirely absorbed into the affine connection of the effective spatial manifold.
Consequently, a fluid particle subjected solely to pressure effects does not experience a force in the Newtonian sense. Rather, it follows a geodesic in a curved space whose geometry is determined by the pressure field.
This interpretation resonates deeply with Einstein's conception of gravity, wherein the gravitational force is likewise an artifact of spacetime curvature. 
The conformal factor $\mathrm{e}^{2\phi}$ provides a vivid geometric visualization: regions of elevated pressure correspond to metric dilation (spatial expansion), whereas low-pressure zones correspond to metric contraction (spatial compression). 
This metric scaling effect, together with the curvature effect encoded in $\tilde{R}_{ij}$, jointly determines the behavior of fluid geodesics, thereby offering a complete geometric description of flow motion.

The framework furnishes a unified and straightforward geometric interpretation of prevailing vortex identification criteria, which have traditionally been formulated on purely kinematic grounds or implicitly embedded with dynamic information. 
The current analysis reveals that these criteria are not heuristic constructs but rather geometric invariants or derived quantities of the effective curved space. 
Specifically, the $Q$ criterion is shown to be proportional to $\tilde{G}$, the $\lambda_{ci}$ criterion emerges as a function of the square root of $\tilde{G}$, the $\lambda_{2}$ criterion is reinterpreted as the condition for the existence of a hyperbolic region in the projected curvature tensor $\tilde{G}_{ij}-\frac{1}{2}\tilde{G}\tilde{h}_{ij}$, and LCS corresponds to the maximal eigen-directions of the Riemann curvature tensor projected onto the geodesic flow.
It is natural to propose a vortex identification criterion based on the pressure-induced tensors, e.g., $\lambda_{\mathrm{sum}}(\tilde{R}_{ij})<0$. 
However, it must be admitted that such manipulation of pressure derivatives would introduce serious error propagation and may not be practical for experimental data.

The degeneration to 2D flows reveals a profound topological constraint that offers a novel interpretation of the dual-cascade scenario in 2D turbulence.
The effective spatial tensor $\tilde{G}_{ij}$ vanishes identically in 2D flows. 
The curvature of the 2D effective space is strictly restricted and is entirely determined by the trace of the Ricci tensor $\tilde{R}^{\mathrm{2D}}\varpropto \nabla^{2}\phi$, allowing only for directionally uniform stretching or contraction.
Remarkably, the integral of the total effective curvature over a closed domain vanishes, $\int \tilde{R}^{\mathrm{2D}}\mathrm{d}^{2}x=0$, as a consequence of the incompressibility constraint and the identity relating enstrophy to strain.
This global topological constraint imposes a strict balance between positive-curvature regions (strain effect dominated) and negative-curvature regions (rotating effect dominated).
Within the current geometric framework, the inverse energy cascade is reinterpreted not as a direct transfer of kinetic energy, but as a geometric condensation of positive-curvature structures toward larger scales, while the forward enstrophy cascade is reinterpreted as a geometric dispersion of negative-curvature structures toward smaller scales.
The coexistence of these two opposing cascades, a hallmark of 2D turbulence, is thus rooted in the fundamental geometric necessity of curvature cancellation.

\begin{acknowledgments}
This work was supported by the Natural Science Foundation of Hubei Province (grant no. 2025AFB187) and the Science Foundation of Wuhan Institute of Technology (grant no. K2023011).
\end{acknowledgments}


\section*{Data Availability Statement}
Data sharing is not applicable to this article as no new data were created or analyzed in this study.

\appendix

\section{Derivation of several fundamental formulas}\label{appA}

\subsection{Expression of the covariant spatial metric}\label{A1}

To prove the expression $\tilde{h}_{ij} = \mathrm{e}^{2\phi} \delta_{ij}$, multiply both sides by $\tilde{h}^{jk}$ and obtain
\begin{equation}
\begin{aligned}
    \tilde{h}_{ij}\tilde{h}^{jk}&=\mathrm{e}^{2\phi} \delta_{ij}\tilde{h}^{jk}\\
    \Rightarrow\delta^{k}_{i}&=\mathrm{e}^{2\phi} \delta_{ij}\mathrm{e}^{-2\phi} \delta^{jk}\\
    \Rightarrow\delta^{k}_{i}&=\delta^{k}_{i}.
    \label{proof_1}
\end{aligned}
\end{equation}
The second step uses the completeness relation (equation ~\ref{con:NC_eqn2_1}), by noting that $v^{i}\tau_{j}=0$.

\subsection{Expressions of spatial connection coefficients (equations~\ref{con:NC_eqn5} and \ref{con:NC_eqn6})}\label{A2}

Given that the metric compatibility condition in NC geometry is $\tilde{\nabla}_{\mu} \tilde{h}^{\nu\rho} = 0$, where $\tilde{\nabla}$ is the covariant derivative with respect to the affine connection, the affine connection $\tilde{\Gamma}^{\rho}_{\mu\nu}$ can be chosen as \cite{hartong_horava-lifshitz_2015}:
\begin{equation}
\begin{split}
    \tilde{\Gamma}_{\mu\nu}^{\rho}=&-\partial^{\rho}\partial_{\mu}\tau_{\nu}+\frac{1}{2}\tilde{h}^{\rho\sigma}\left(\partial_{\mu}\tilde{h}_{\nu\sigma}+\partial_{\nu}\tilde{h}_{\mu\sigma}-\partial_{\sigma}\tilde{h}_{\mu\nu}\right)\\&+\tilde{h}^{\rho\sigma}\tau_{(\mu}F_{\nu)\sigma}.
    \label{proof_2}
\end{split}
\end{equation}
Recall that the curvature occurs only in space and the absoluteness of time (i.e., $\partial_{\mu}\tau_{\nu}=\partial_{\mu}\partial_{\nu}t=0$), the space-time term vanishes:
\begin{equation}
\partial^{\rho}\partial_{\mu}\tau_{\nu}=\partial^{\rho}0=0.
    \label{proof_2.1}
\end{equation}
Therefore, the spacetime indices are replaced by space indices and the spatial part of the above equation is transformed into equation~\ref{con:NC_eqn5},
\begin{equation}
\tilde{\Gamma}_{jk}^{i}=\frac{1}{2}\tilde{h}^{il}\left(\partial_{j}\tilde{h}_{kl}+\partial_{k}\tilde{h}_{jl}-\partial_{l}\tilde{h}_{jk}\right).
    \label{proof_3}
\end{equation}
Substituting $\tilde{h}_{ij} = \mathrm{e}^{2\phi} \delta_{ij}$ into $\frac{1}{2}\tilde{h}^{il}\partial_{j}\tilde{h}_{kl}$ yields
\begin{equation}
\begin{aligned}
    \frac{1}{2}\tilde{h}^{il}\partial_{j}\tilde{h}_{kl}&=\frac{1}{2}\tilde{h}^{il}\partial_{j}\left(\mathrm{e}^{2\phi} \delta_{kl}\right)\\
    &=\frac{1}{2}\tilde{h}^{il}\left(\delta_{kl}\partial_{j}\mathrm{e}^{2\phi}+\mathrm{e}^{2\phi}\partial_{j}\delta_{kl}\right)\\
    &=\frac{1}{2}\left(\mathrm{e}^{-2\phi} \delta^{il}\right)\left(2\mathrm{e}^{2\phi}\delta_{kl}\partial_{j}\phi+0\right)\\
    &=\delta^{i}_{k}\partial_{j}\phi.
    \label{proof_4}
\end{aligned}
\end{equation}
Similarly, substituting $\tilde{h}_{ij} = \mathrm{e}^{2\phi} \delta_{ij}$ into $\frac{1}{2}\tilde{h}^{il}\partial_{k}\tilde{h}_{jl}$ and $\frac{1}{2}\tilde{h}^{il}\partial_{l}\tilde{h}_{jk}$ yields $\delta^{i}_{j}\partial_{k}\phi$ and $\delta^{il}\delta_{jk}\partial_{j}\phi$, respectively. Note that $\partial^i \phi = \delta^{il} \partial_l \phi$, $\delta^{il}\delta_{jk}\partial_{j}\phi$ is equivalent to $\delta_{jk}\partial^{i}\phi$. Equation~\ref{con:NC_eqn6} is proved.

$F_{\mu\nu}$ is an arbitrary two‑form encoding the observer degrees of freedom, which is at our disposal.
To eliminate the term $\tilde{\Gamma}^{i}_{0j}u^{j}$ from the geodesic equation and to ensure that the observer's spatial axes are non‑rotating relative to the fluid, we set
\begin{equation}
F_{jk}=0 \Rightarrow \tilde{\Gamma}^{i}_{0j}=0. 
    \label{proof_4.1}
\end{equation}
The time‑time component $\Gamma^{i}_{00}$ is the crucial element that will encode the pressure gradient. A natural choice guided by the analogy with the Newtonian limit of general relativity is
\begin{equation}
F_{0i}=-2\partial_{i}\phi. 
    \label{proof_4.2}
\end{equation}
Then the time‑time component is derived as:
\begin{equation}
\tilde{\Gamma}^{i}_{00}=\tilde{h}^{ij}\frac{1}{2}F_{0j}=\tilde{h}^{ij}\partial_{j}\phi. 
    \label{proof_4.3}
\end{equation}
To this extent, the complete Newton–Cartan connection is concluded in table~\ref{Tab:A1}.

\begin{table}
	\begin{center}
		\def~{\hphantom{0}}
        \begin{threeparttable}
		\begin{tabular}{cc}
        \toprule
			Component & Expression\\[1pt]
            \hline
            time–time component $\tilde{\Gamma}^{i}_{00}$ & $\tilde{h}^{ij}\partial_{j}\phi$\\
			time–space component  $\tilde{\Gamma}^{i}_{0j}$ & 0 \\
            pure space component $\tilde{\Gamma}^{i}_{jk}$ & $\delta_j^i\partial_k\phi+\delta_k^i\partial_j\phi-\delta_{jk}\partial^i\phi$ \\
            \toprule
		\end{tabular}
        \end{threeparttable}
		\caption{Summary of the connection coefficients.}
		\label{Tab:A1}
	\end{center}
\end{table}

\subsection{Computation of the spatial tensors and corresponding traces (equations~\ref{con:NC_eqn7} to \ref{con:NC_eqn9})}\label{A3}
The spatial Ricci tensor is defined as:
\begin{equation}
\tilde{R}_{ij}=\partial_k\tilde{\Gamma}_{ij}^k-\partial_i\tilde{\Gamma}_{kj}^k+\tilde{\Gamma}_{kl}^k\tilde{\Gamma}_{ij}^l-\tilde{\Gamma}_{il}^k\tilde{\Gamma}_{kj}^l.
    \label{proof_5}
\end{equation}
The first term $\partial_k\tilde{\Gamma}_{ij}^k$ can be further derived as:
\begin{equation}
\begin{aligned}
    \partial_k\tilde{\Gamma}_{ij}^k&=\partial_k\left(\delta^k_i\partial_j\phi+\delta^k_j\partial_i\phi-\delta_{ij}\partial^{k}\phi\right)\\
    &=\partial_j\phi\partial_k\delta^k_i+\delta^k_i\partial_k\partial_j\phi+\partial_i\phi\partial_k\delta^k_j\\&+\delta^k_j\partial_k\partial_i\phi-\partial_k\delta_{ij}\partial^k\phi-\delta_{ij}\partial_k\partial^k\phi\\
    &=\partial_i\partial_j\phi+\partial_j\partial_i\phi-\delta_{ij}\nabla^{2}\phi\\
    &=2\partial_i\partial_j\phi-\delta_{ij}\nabla^{2}\phi.
    \label{proof_6}
\end{aligned}
\end{equation}
The second term $\partial_i\tilde{\Gamma}_{kj}^k$ can be derived as:
\begin{equation}
\begin{aligned}
    \partial_i\tilde{\Gamma}_{kj}^k&=\partial_i\left(\delta^k_k\partial_j\phi+\delta^k_j\partial_k\phi-\delta_{kj}\partial^{k}\phi\right)\\
    &=\partial_j\phi\partial_i\delta^k_k+\delta^k_k\partial_i\partial_j\phi+\partial_k\phi\partial_i\delta^k_j\\
    &+\delta^k_j\partial_i\partial_k\phi-\partial_i\delta_{kj}\partial^k\phi-\delta_{kj}\partial_i\partial^k\phi\\
    &=3\partial_i\partial_j\phi+\partial_i\partial_j\phi-\partial_i\partial_j\phi\\
    &=3\partial_i\partial_j\phi.
    \label{proof_7}
\end{aligned}
\end{equation}
The third term $\tilde{\Gamma}_{kl}^k\tilde{\Gamma}_{ij}^l$ can be derived as:
\begin{equation}
\begin{aligned}
    \tilde{\Gamma}_{kl}^k\tilde{\Gamma}_{ij}^l&=\left(\delta^k_k\partial_l\phi+\delta^k_l\partial_k\phi-\delta_{kl}\partial^k\phi\right)\\
    &\cdot\left(\delta^l_i\partial_j\phi+\delta^l_j\partial_i\phi-\delta_{ij}\partial^l\phi\right)\\
    &=3\partial_l\phi\left(\delta^l_i\partial_j\phi+\delta^l_j\partial_i\phi-\delta_{ij}\partial^l\phi\right)\\
    &=3\partial_i\phi\partial_j\phi+3\partial_j\phi\partial_i\phi-3\delta_{ij}\left(\nabla\phi\right)^2\\
    &=6\partial_i\phi\partial_j\phi-3\delta_{ij}\left(\nabla\phi\right)^2.
    \label{proof_8}
\end{aligned}
\end{equation}
The fourth term $\tilde{\Gamma}_{il}^k\tilde{\Gamma}_{kj}^l$ can be derived as:
\begin{equation}
\begin{aligned}
    \tilde{\Gamma}_{il}^k\tilde{\Gamma}_{kj}^l&=\left(\delta^k_i\partial_l\phi+\delta^k_l\partial_i\phi-\delta_{il}\partial^k\phi\right)\\
    &\cdot\left(\delta^l_k\partial_j\phi+\delta^l_j\partial_k\phi-\delta_{kj}\partial^l\phi\right)\\
    &=\left(\delta^k_i\delta^l_k\partial_l\phi\partial_j\phi+\delta^k_i\delta^l_j\partial_l\phi\partial_k\phi-\delta_{ij}\partial_l\phi\partial^l\phi\right)\\
    &+(\delta^k_l\delta^l_k\partial_i\phi\partial_j\phi+\delta^k_l\delta^l_j\partial_i\phi\partial_k\phi-\delta_{lj}\partial_i\phi\partial^l\phi)\\
    &-\left(\delta_{il}\delta^l_k\partial^k\phi\partial_j\phi+\delta_{il}\delta^l_j\partial^k\phi\partial_k\phi-\delta_{il}\delta_{kj}\partial^k\phi\partial^l\phi\right)\\
    &=\left[\partial_i\phi\partial_j\phi+\partial_j\phi\partial_i\phi-\delta_{ij}(\nabla\phi)^2\right]\\
    &+(3\partial_i\phi\partial_j\phi+\partial_j\phi\partial_i\phi-\partial_j\phi\partial_i\phi)\\
    &-\left[\partial_i\phi\partial_j\phi+\delta_{ij}(\nabla\phi)^2-\partial_j\phi\partial_i\phi\right]\\
    &=5\partial_i\phi\partial_j\phi-2\delta_{ij}(\nabla\phi)^2.
    \label{proof_9}
\end{aligned}
\end{equation}
Substituting equations~\ref{proof_6}-\ref{proof_9} back into equation~\ref{proof_5}, we then obtain equation~\ref{con:NC_eqn7}.
And the trace of $\tilde{R}_{ij}$ (equation~\ref{con:NC_eqn8}) can be obtained as follows:
\begin{equation}
\begin{aligned}
    \tilde{R}&=\tilde{h}^{ij}\tilde{R}_{ij}=\mathrm{e}^{-2\phi}\delta^{ij}\tilde{R}_{ij}\\
    &=\mathrm{e}^{-2\phi}\delta^{ij}\left[-\partial_i\partial_j\phi-\delta_{ij}\nabla^2\phi+\partial_i\phi\partial_j\phi-\delta_{ij}(\nabla\phi)^2\right]\\
    &=\mathrm{e}^{-2\phi}\left[-\partial^j\partial_j\phi-3\nabla^2\phi+\partial^j\phi\partial_j\phi-3(\nabla\phi)^2\right]\\
    &=\mathrm{e}^{-2\phi}\left[-4\nabla^2\phi-2(\nabla\phi)^2\right].
    \label{proof_10}
\end{aligned}
\end{equation}

The spatial tensor $\tilde{G}_{ij}$ and its trace are therefore calculated as:
\begin{equation}
\begin{aligned}
    \tilde{G}_{ij}&=\tilde{R}_{ij}-\frac{1}{2}\tilde{R}\tilde{h}_{ij}\\
    &=-\partial_i\partial_j\phi-\delta_{ij}\nabla^2\phi+\partial_i\phi\partial_j\phi-\delta_{ij}(\nabla\phi)^2\\
    &-\frac{1}{2}\mathrm{e}^{-2\phi}\left[-4\nabla^2\phi-2(\nabla\phi)^2\right]\mathrm{e}^{2\phi}\delta_{ij}\\
    &=-\partial_i\partial_j\phi-\delta_{ij}\nabla^2\phi+\partial_i\phi\partial_j\phi-\delta_{ij}(\nabla\phi)^2\\
    &+2\delta_{ij}\nabla^2\phi+\delta_{ij}(\nabla\phi)^2\\
    &=-\partial_i\partial_j\phi+\delta_{ij}\nabla^2\phi+\partial_i\phi\partial_j\phi,
    \label{proof_11}
\end{aligned}
\end{equation}
\begin{equation}
\begin{aligned}
    \tilde{G}&=\tilde{h}^{ij}\tilde{G}_{ij}=\mathrm{e}^{-2\phi}\delta^{ij}\tilde{G}_{ij}\\
    &=\mathrm{e}^{-2\phi}\delta^{ij}\left(-\partial_i\partial_j\phi+\delta_{ij}\nabla^2\phi+\partial_i\phi\partial_j\phi\right)\\
    &=\mathrm{e}^{-2\phi}\left(-\partial^j\partial_j\phi+3\nabla^2\phi+\partial^j\phi\partial_j\phi\right)\\
    &=\mathrm{e}^{-2\phi}\left[2\nabla^2\phi+(\nabla\phi)^2\right].
    \label{proof_12}
\end{aligned}
\end{equation}

It should be particularly noted that when degenerate to two dimensions, the 2D Ricci tensor $\tilde{R}^\mathrm{2D}_{ij}$ and 2D spatial tensor $\tilde{G}^\mathrm{2D}_{ij}$ undergo significant changes since $\delta^i_i=2$ instead of 3.
To obtain these updated tensors, the four terms on the right hand side of equation~\ref{proof_5} are updated as follows.

\begin{equation}
\begin{aligned}
    \partial_i\tilde{\Gamma}_{kj}^k&=\partial_i\left(\delta^k_k\partial_j\phi+\delta^k_j\partial_k\phi-\delta_{kj}\partial^{k}\phi\right)\\
    &=\partial_j\phi\partial_i\delta^k_k+\delta^k_k\partial_i\partial_j\phi+\partial_k\phi\partial_i\delta^k_j\\
    &+\delta^k_j\partial_i\partial_k\phi-\partial_i\delta_{kj}\partial^k\phi-\delta_{kj}\partial_i\partial^k\phi\\
    &=2\partial_i\partial_j\phi+\partial_i\partial_j\phi-\partial_i\partial_j\phi\\
    &=2\partial_i\partial_j\phi.
    \label{proof_7_1}
\end{aligned}
\end{equation}

\begin{equation}
\begin{aligned}
    \tilde{\Gamma}_{kl}^k\tilde{\Gamma}_{ij}^l&=\left(\delta^k_k\partial_l\phi+\delta^k_l\partial_k\phi-\delta_{kl}\partial^k\phi\right)\\
    &\cdot\left(\delta^l_i\partial_j\phi+\delta^l_j\partial_i\phi-\delta_{ij}\partial^l\phi\right)\\
    &=2\partial_l\phi\left(\delta^l_i\partial_j\phi+\delta^l_j\partial_i\phi-\delta_{ij}\partial^l\phi\right)\\
    &=2\partial_i\phi\partial_j\phi+2\partial_j\phi\partial_i\phi-2\delta_{ij}\left(\nabla\phi\right)^2\\
    &=4\partial_i\phi\partial_j\phi-2\delta_{ij}\left(\nabla\phi\right)^2.
    \label{proof_8_1}
\end{aligned}
\end{equation}

\begin{equation}
\begin{aligned}
    \tilde{\Gamma}_{il}^k\tilde{\Gamma}_{kj}^l&=\left(\delta^k_i\partial_l\phi+\delta^k_l\partial_i\phi-\delta_{il}\partial^k\phi\right)\\
    &\cdot\left(\delta^l_k\partial_j\phi+\delta^l_j\partial_k\phi-\delta_{kj}\partial^l\phi\right)\\
    &=\left(\delta^k_i\delta^l_k\partial_l\phi\partial_j\phi+\delta^k_i\delta^l_j\partial_l\phi\partial_k\phi-\delta_{ij}\partial_l\phi\partial^l\phi\right)\\
    &+(\delta^k_l\delta^l_k\partial_i\phi\partial_j\phi+\delta^k_l\delta^l_j\partial_i\phi\partial_k\phi-\delta_{lj}\partial_i\phi\partial^l\phi)\\
    &-\left(\delta_{il}\delta^l_k\partial^k\phi\partial_j\phi+\delta_{il}\delta^l_j\partial^k\phi\partial_k\phi-\delta_{il}\delta_{kj}\partial^k\phi\partial^l\phi\right)\\
    &=\left[\partial_i\phi\partial_j\phi+\partial_j\phi\partial_i\phi-\delta_{ij}(\nabla\phi)^2\right]\\
    &+(2\partial_i\phi\partial_j\phi+\partial_j\phi\partial_i\phi-\partial_j\phi\partial_i\phi)\\
    &-\left[\partial_i\phi\partial_j\phi+\delta_{ij}(\nabla\phi)^2-\partial_j\phi\partial_i\phi\right]\\
    &=4\partial_i\phi\partial_j\phi-2\delta_{ij}(\nabla\phi)^2.
    \label{proof_9_2}
\end{aligned}
\end{equation}

Then the 2D Ricci tensor and its trace are given as:
\begin{equation}
\begin{aligned}
\tilde{R}_{ij}^\mathrm{2D}&=2\partial_i\partial_j\phi-\delta_{ij}\nabla^2\phi-2\partial_i\partial_j\phi+4\partial_i\phi\partial_j\phi\\
&-2\delta_{ij}(\nabla\phi)^2-4\partial_i\phi\partial_j\phi+2\delta_{ij}(\nabla\phi)^2\\
&=-\delta_{ij}\nabla^2\phi.
    \label{proof_16}
\end{aligned}
\end{equation}
\begin{equation}
\begin{aligned}
\tilde{R}^\mathrm{2D}&=\tilde{h}^{ij}\tilde{R}^\mathrm{2D}_{ij}=\mathrm{e}^{-2\phi}\delta^{ij}\tilde{R}_{ij}^\mathrm{2D}\\
&=-\mathrm{e}^{-2\phi}\delta^{ij}\delta_{ij}\nabla^2\phi\\
&=-2\mathrm{e}^{-2\phi}\nabla^2\phi.
    \label{proof_17}
\end{aligned}
\end{equation}
The corresponding 2D spatial tensor $\tilde{G}^\mathrm{2D}_{ij}$ then follows as: 
\begin{equation}
\begin{aligned}
\tilde{G}^\mathrm{2D}_{ij}&=\tilde{R}^\mathrm{2D}_{ij}-\frac{1}{2}\tilde{R}^\mathrm{2D}\tilde{h}_{ij}\\
&=-\delta_{ij}\nabla^2\phi-\frac{1}{2}(-2\mathrm{e}^{-2\phi}\nabla^2\phi)\mathrm{e}^{2\phi}\delta_{ij}\\
&=0.
    \label{proof_18}
\end{aligned}
\end{equation}
Consequently, the trace $\tilde{G}^\mathrm{2D}$ is zero. 
For a two-dimensional manifold, the Ricci tensor completely determines the curvature, and the spatial tensor vanishes automatically.




\subsection{Computation of the projection of the pressure Hessian (equation~\ref{c2_2})}\label{A3.5}

Substituting the expressions of $\tilde{G}_{ij}$ and the corresponding trace $\tilde{G}$ into equation~\ref{c2_1} in the weak-field limit, the symmetric tensor can be projected as:
\begin{equation}
\begin{aligned}
\boldsymbol{S}^{2}+\boldsymbol{\Omega}^{2}&=-\partial^{2}_{ij}\phi=\tilde{G}_{ij}-\delta_{ij}\nabla^{2}\phi-\partial_{i}\phi\partial_{j}\phi\\
&\simeq\tilde{G}_{ij}-\delta_{ij}\nabla^{2}\phi=\tilde{G}_{ij}-\delta_{ij}\left(\frac{\tilde{G}}{2}\right)\\
&\simeq\tilde{G}_{ij}-\frac{1}{2}\tilde{G}\tilde{h}_{ij}.
    \label{proof_222}
\end{aligned}
\end{equation}
Similarly, the projection of the symmetric tensor onto the plane spanned by $\tilde{R}_{ij}$ and its trace $\tilde{R}$ can be obtained as:
\begin{equation}
\begin{aligned}
\boldsymbol{S}^{2}+\boldsymbol{\Omega}^{2}&=-\partial^{2}_{ij}\phi=\tilde{R}_{ij}+\delta_{ij}\nabla^{2}\phi-\partial_{i}\phi\partial_{j}\phi+\delta_{ij}(\nabla\phi)^2\\
&\simeq\tilde{R}_{ij}+\delta_{ij}\nabla^{2}\phi=\tilde{R}_{ij}+\delta_{ij}\left(-\frac{\tilde{R}}{4}\right)\\
&\simeq\tilde{R}_{ij}-\frac{1}{4}\tilde{R}\tilde{h}_{ij}.
    \label{proof_223}
\end{aligned}
\end{equation}

\subsection{Computation of the Riemann curvature tensor (equation~\ref{lcs_2})}\label{A4}
Since the Weyl tensor vanishes identically, the Riemann curvature tensor in three dimensions is completely determined by the Ricci tensor and the scalar curvature:
\begin{equation}
\tilde{R}_{ijkl}=\frac{\tilde{R}}{2}(\tilde{h}_{il}\tilde{h}_{jk}-\tilde{h}_{ik}\tilde{h}_{jl})-\tilde{h}_{il}\tilde{R}_{jk}+\tilde{h}_{ik}\tilde{R}_{jl}-\tilde{h}_{jk}\tilde{R}_{il}+\tilde{h}_{jl}\tilde{R}_{ik}
    \label{proof_19}
\end{equation}
Substituting the expressions of $\tilde{R}_{ij}$, $\tilde{R}$ and $\tilde{h}_{ij}$ into equation~\ref{proof_19}, we can obtain the Riemann curvature tensor in the weak-field limit:
\begin{equation}
\begin{aligned}
\tilde{R}_{ijkl}&\simeq \frac{-4\nabla^{2}\phi}{2}(\delta_{il}\delta_{jk}-\delta_{ik}\delta_{jl})-\delta_{il}(-\partial_{j}\partial_{k}\phi-\delta_{jk}\nabla^{2}\phi)\\
&+\delta_{ik}(-\partial_{j}\partial_{l}\phi-\delta_{jl}\nabla^{2}\phi)-\delta_{jk}(-\partial_{i}\partial_{l}\phi-\delta_{il}\nabla^{2}\phi)\\
&+\delta_{jl}(-\partial_{i}\partial_{k}\phi-\delta_{ik}\nabla^{2}\phi)\\
&=-\delta_{il}\partial_{j}\partial_{k}\phi+\delta_{ik}\partial_{j}\partial_{l}\phi-\delta_{jk}\partial_{i}\partial_{l}\phi+\delta_{jl}\partial_{i}\partial_{k}\phi.
    \label{proof_20}
\end{aligned}
\end{equation}

\nocite{*}
\bibliography{pressure}

\end{document}